\documentclass[amsmath,amssymb,twocolumn,showpacs,aps,prb,showkeys, 10pt]{revtex4-2} 

\usepackage{graphicx}
\usepackage{ifthen}
\usepackage{color}
\usepackage{bm}
\usepackage{braket}
\usepackage{multirow}
\usepackage{hyperref}
\usepackage{soul}
\usepackage{subfigure}
\hypersetup{%backref=true,
 pdfnewwindow=true, colorlinks=true,
 linkcolor=blue, anchorcolor=blue,
 citecolor=blue, filecolor=blue,
 menucolor=blue, urlcolor=blue}
\usepackage{listings}
\begin{document}

\title{Stability-superconductivity map for compressed Na-intercalated graphite}

\author{Shashi B. Mishra}
\email{mshashi125@gmail.com}
\affiliation{Department of Physics, and Astronomy, Binghamton University-SUNY, Binghamton, New York 13902, USA}
\author{Edan T. Marcial}
\affiliation{Department of Physics, and Astronomy, Binghamton University-SUNY, Binghamton, New York 13902, USA}
\author{Suryakanti Debata}
\affiliation{Department of Physics, and Astronomy, Binghamton University-SUNY, Binghamton, New York 13902, USA}
\author{Aleksey N. Kolmogorov }
\email{kolmogorov@binghamton.edu}
\affiliation{Department of Physics, and Astronomy, Binghamton University-SUNY, Binghamton, New York 13902, USA}
\author{Elena R. Margine}
\email{rmargine@binghamton.edu}
\affiliation{Department of Physics, and Astronomy, Binghamton University-SUNY, Binghamton, New York 13902, USA}
\date{\today}
%\keywords{graphite intercalation compounds, thermodynamic stability, superconductivity, electron-phonon coupling.}

\begin{abstract}
A recent {\it ab initio} investigation of Na-C binary compounds under moderate pressures has uncovered a possible stable NaC$_4$ superconductor with an estimated critical temperature up to 41~K. We revisit this promising binary system by performing a more focused exploration of Na-intercalated graphite configurations, assessing the sensitivity of their thermodynamic stability to density functional approximations at different $(T, P)$ conditions, and examining their superconducting properties with the Migdal-Eliashberg formalism. The combinatorial screening of possible Na arrangements reveals additional stable stoichiometries, {\it i.e.}, Na$_3$C$_{10}$, NaC$_8$, NaC$_{10}$, and NaC$_{12}$, that redefine the previously proposed convex hulls for pressures up to 10~GPa. The evaluation of formation enthalpies with different van der Waals functionals indicates that the proposed compounds might not be thermodynamically stable at zero temperature but some of them could stabilize due to the vibrational entropy or form via cold compression if graphite is used as a starting material. Our more rigorous modeling of the electron-phonon coupling in NaC$_4$ confirms the material’s potential for high-temperature superconductivity, with a critical temperature reaching 48~K at 10~GPa, and reveals a well-defined two-gap structure unusual for an electron-doped compound. By tracking the position of the intercalant nearly free electron states with respect to the Fermi level in viable Na-C compounds, we map out the range of pressures and compositions needed for strong electron-phonon coupling and identify Na$_3$C$_{10}$ as an equally promising superconductor.
\end{abstract}

\maketitle

\section{\label{sec:intro}Introduction}

Carbon-based materials offer a rich playground in the quest for thermodynamically stable superconductors at ambient and low pressure~\cite{Jishi1992,Ekimov2004,Okazaki2015,Heguri2015,Zhou2021,Bhaumik2017,Emery2008}. Graphite intercalation compounds (GICs)~\cite{Dresselhaus1981}, in particular, have been known to superconduct since the mid 1960s when superconductivity below 1~K was discovered in graphite intercalated with alkali metals (AC$_8$ with A = K, Rb, and Cs)~\cite{Hanay1965}. Since then, considerable efforts have been made to find superconductivity in GICs at elevated temperatures, and the search has been focused on increasing the charge transfer from the intercalant atoms to the host graphene sheets by varying the stoichiometry and the intercalant type. To date, YbC$_6$ and CaC$_6$ with critical temperature ($T_{\rm c}$) values of 6.5~K and 11.5~K stand as the highest temperature superconductors in this class under ambient conditions~\cite{Weller2005,Emery2005}. Superconductivity has also been induced or enhanced by applying external pressure. For instance, LiC$_2$ and NaC$_2$ become superconductors with $T_{\rm c}=1.9$~K at 3.3~GPa~\cite{Belash1989} and $T_{\rm c}=5.0$~K at 3.5~GPa~\cite{Belash1987}, while in CaC$_6$ a maximum value of 15.1~K is reached at 7.5~GPa~\cite{Gauzzi2007}. 

\begin{figure*}[!hbt]
    \centering
    \includegraphics[width=17.5cm]{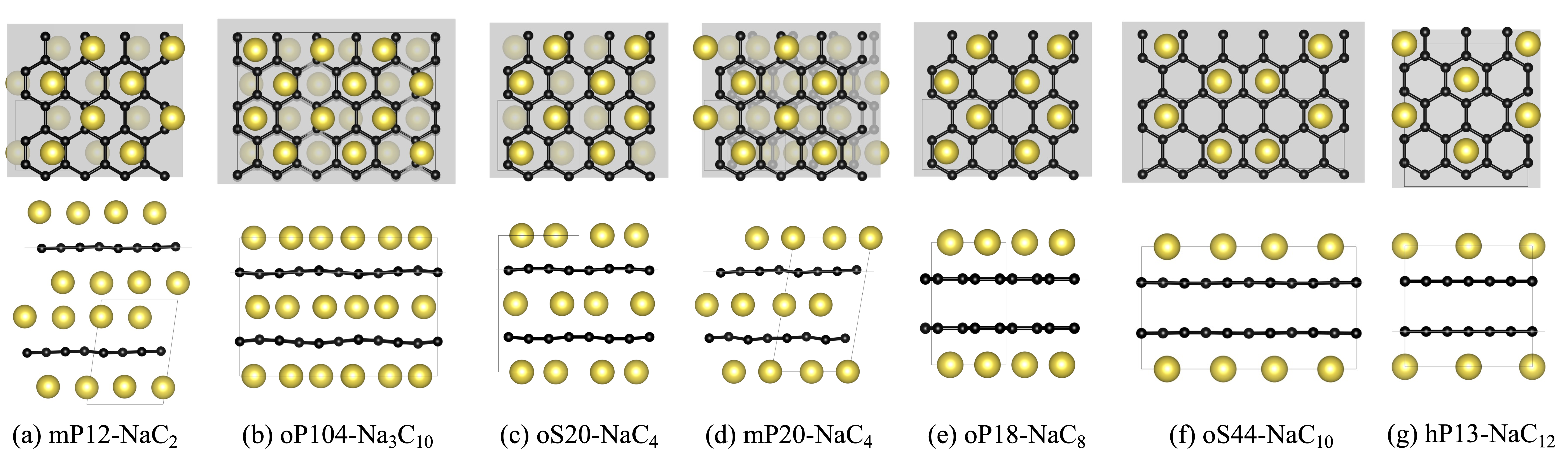}
    \caption{Select high-pressure Na-C phases found to be viable in previous~\cite{Hao2023} (a,c) and present (b,d,e,f,g) {\it ab initio} studies. The top and side views show the corresponding distributions of Na (large yellow spheres) intercalated between C honeycomb layers (small black spheres). The figures were generated with VESTA~\cite{Momma2011}.}
    \label{fig:struct}
\end{figure*}

Recently, an \textit{ab initio} evolutionary structure search in the Na-C system has identified a new GIC at the 1:4 composition~\cite{Hao2023}. The NaC$_4$ phase with the oS20 Pearson symbol and \textit{Cmcm} symmetry (Fig.~\ref{fig:struct}(c)) was predicted to stabilize above 8.9~GPa and have a maximum $T_{\rm c}$ of 41.2~K at 5~GPa. While this work has shown that the electron-phonon (e-ph) coupling is strong enough to yield a $T_{\rm c}$ above the record 39~K value in MgB$_2$ based on the Allen-Dynes modified McMillan formula~\cite{McMillan1968,Allen1975}, the nature of the superconducting gap has not been investigated.

Our present study reexamines the stability and superconductivity of Na-C candidate materials under pressures up to 10~GPa. Further exploration of the vast configuration space was motivated by the well-known difficulty of locating thermodynamically stable compounds even with the most advanced structure optimization methods ~\cite{Oganov2011,Oganov2019,Wang2014,ak45}. For instance, our unconstrained evolutionary searches yielded a number of new complex crystal structures that have either been confirmed experimentally~\cite{ak16,ak23,ak41} or inspired the creation of more stable derivatives based on the identified favorable motifs~\cite{ak24,ak47,ak49,Charlsey2024,ak52}. In the latest investigation of layered metal borocarbides, we generated viable quaternary phases via systematic screening of possible metal decorations within ideal or modified honeycomb frameworks~\cite{ak52}. Application of this strategy to Na-C in the present work resulted in a significant revision of the previously proposed ground states. The updated set at 10~GPa contains large-sized oP104-Na$_3$C$_{10}$, oP18-NaC$_8$, oS44-NaC$_{10}$, and hP13-NaC$_{12}$ phases that make mP12-NaC$_2$ metastable and oS20-NaC$_4$ marginally stable. 

Importantly, we demonstrate that the global stability of these van der Waals solids is very sensitive to the treatment of the dispersive interactions. While the Na-C are thermodynamically stable in the optB88-vdW approximation of the non-local density functional~\cite{Klimes2010}, as was shown in Ref.~\cite{Hao2023}, they have positive formation enthalpies in our optB86b-vdW~\cite{Klime2011} calculations. Fortunately, the inclusion of vibrational entropy brings them closer to stability and makes Na$_3$C$_{10}$ a true ground state at room temperature. Moreover, we argue that the feasibility of obtaining the binary layered intercalated compounds should be evaluated with respect to graphite rather than diamond, provided that the synthesis starts with the former material and the $(T,P)$ conditions are kept sufficiently low to avoid the $sp^2$ to $sp^3$ transformation. When referenced to the layered carbon polymorph, the Na-intercalated variants have negative formation Gibbs free energies. 

Finally, we probe the proposed materials for superconducting properties. We analyze the position of the intercalant nearly free electron states across different phases and pressures to establish conditions needed for strong e-ph coupling. According to the constructed composition-pressure map, Na$_3$C$_{10}$ compound with a complex structure has promise to be a high-$T_{\rm c}$ superconductor. We further study oS20-NaC$_4$ at 10~GPa using the fully anisotropic Migdal-Eliashberg (aME) formalism~\cite{Margine2013,Lee2023,Lucrezi2024}. The aME method has been demonstrated to yield significantly higher estimates of the critical temperature, by a factor of two to three, in related layered conventional superconductors, {\it e.g.}, metal borides and borocarbides~\cite{Liu2001,Choi2002,Choi2009,Margine2013,Kafle2022,Wang2023,Charlsey2024,tomassetti2024}. We calculate the momentum dependence of the e-ph coupling and superconducting energy gap on the Fermi surface. Our aME calculations yield a $T_{\rm c}$ of 48~K, which is slightly larger than the estimate of 39~K obtained in the previous work using an isotropic treatment of the e-ph coupling~\cite{Hao2023}.

\section{\label{sec:methods}Methods}

We examined the stability of the Na-C phases with VASP~\cite{Kresse1996} using projector augmented wave potentials~\cite{Blochl1994}. By default, the non-local van der Waals interactions were treated with the optB86b-vdW functional~\cite{Klime2011}, but also checked with alternative optB88-vdW~\cite{Klimes2010} and r$^2$SCAN+rVV10~\cite{Furness2020,Ning2022} functionals. Dense Monckhorst-Pack $k$-point meshes with $\Delta k\sim 2\pi\times 0.025$ \AA$^{-1}$ and a plane-wave cutoff of 500~eV ensured numerical convergence to typically within 1 meV/atom. To systematically screen possible decorations of interlayer sites, we constructed supercells with up to 42 atoms by expanding the stoichiometric MgB$_2$ prototype and sequentially removing metal atoms. For detection and elimination of equivalent configurations, we relied on our radial distribution function fingerprint~\cite{ak41}. We calculated thermodynamic corrections due to the vibrational entropy using the finite displacement method implemented in {\small PHONOPY}~\cite{Togo2015}, employing supercells sized between 52 and 128 atoms and applying 0.04~\AA{} displacements within the harmonic approximation. Full structural information for select phases is provided in the form of CIF files in Supplemental Material~\cite{SM}.

For calculating properties related to superconductivity, we employed the Quantum {\small ESPRESSO} (QE) package~\cite{Giannozzi2017} with the optB86b-vdW functional~\cite{Klime2011,Thonhauser2015,Berland2015} and optimized norm-conserving Vanderbilt pseudopotentials ({\small ONCVPSP})~\cite{Hamann2013} from the Pseudo Dojo library~\cite{Vansetten2018} generated with the relativistic Perdew-Burke-Ernzerhof parametrization~\cite{Perdew1996}. We used a plane-wave cutoff of 100~Ry, a Methfessel-Paxton smearing~\cite{Methfessel1989} value of 0.02~Ry, and $\Gamma$-centered $k$-grids of $12 \times 12 \times 8$ for mP12-NaC$_{2}$, $12 \times 12 \times 8$ for oS52-Na$_3$C$_{10}$, $6 \times 6 \times 4$ for oP104-Na$_3$C$_{10}$, $16 \times 16 \times 16$ for oS20-NaC$_4$, $12 \times 12 \times 6$ for mP20-NaC$_4$, $6 \times 4 \times 5$ for oP18-NaC$_8$, $12 \times 12 \times 12$ for oS44-NaC$_{10}$, and $12 \times 12 \times 8$ for mP12-NaC$_{12}$ to describe the electronic structure. The lattice parameters and atomic positions were relaxed until the total enthalpy was converged within $10^{-6}$~Ry and the maximum force on each atom was less than $10^{-4}$ Ry/\AA. The dynamical matrices and the linear variation of the self-consistent potential were calculated within density-functional perturbation theory~\cite{Baroni2001} on $q$-meshes of $3 \times 3 \times 2$ for mP12-NaC$_{2}$, $3 \times 3 \times 2$ for oS52-Na$_3$C$_{10}$, $1 \times 1 \times 1$ for oP104-Na$_3$C$_{10}$, $4 \times 4 \times 4$ for oS20-NaC$_4$, $4 \times 4 \times 2$ for mP20-NaC$_4$, $3 \times 2 \times 2$ for oP18-NaC$_8$, $3 \times 3 \times 3$ for oS44-NaC$_{10}$, and $4 \times 4 \times 3$ for mP12-NaC$_{12}$.

\begin{figure*}
    \centering
    \subfigure
    {\includegraphics[width=0.49\textwidth]{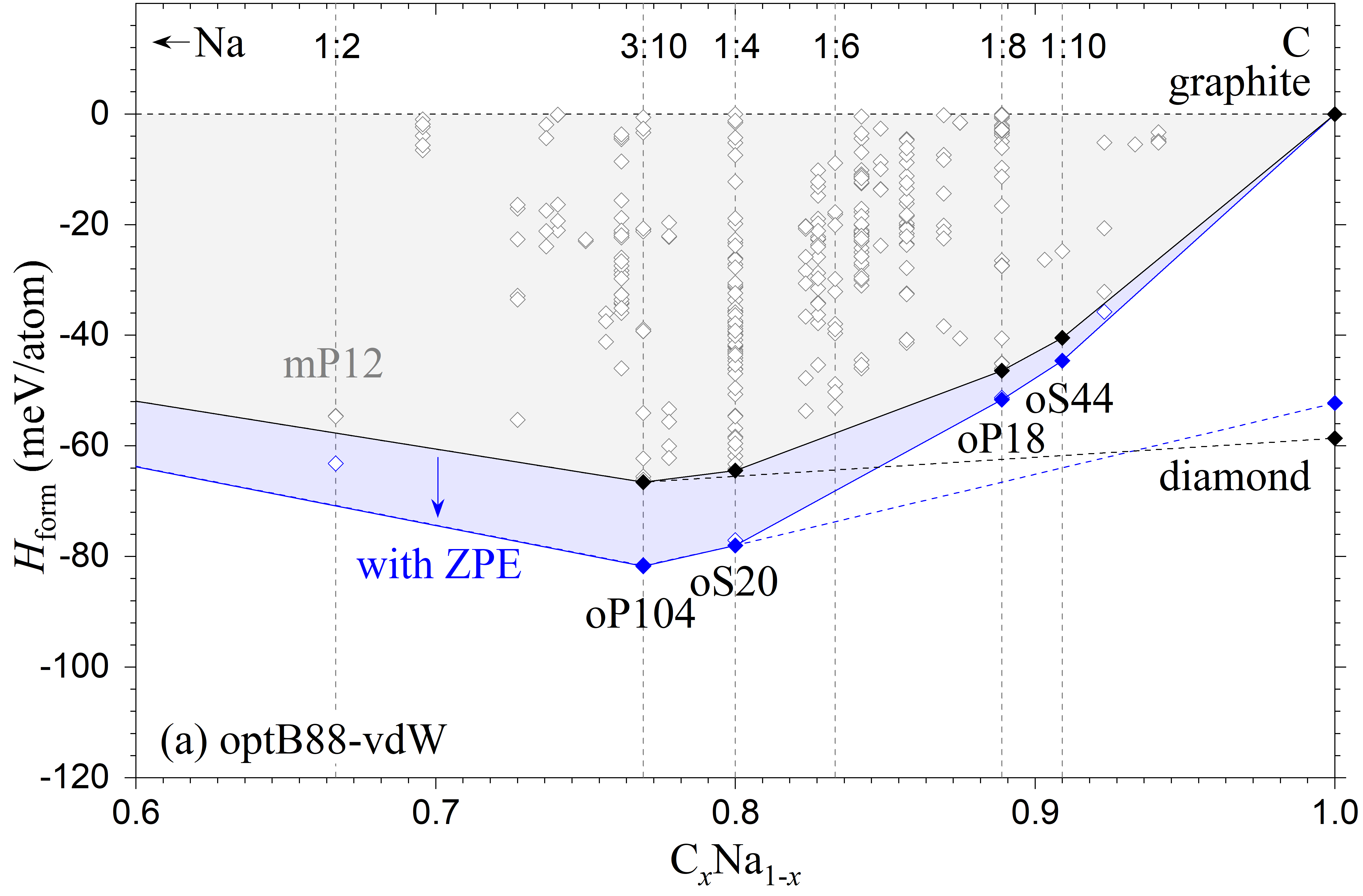}}
    \subfigure
    {\includegraphics[width=0.49\textwidth]{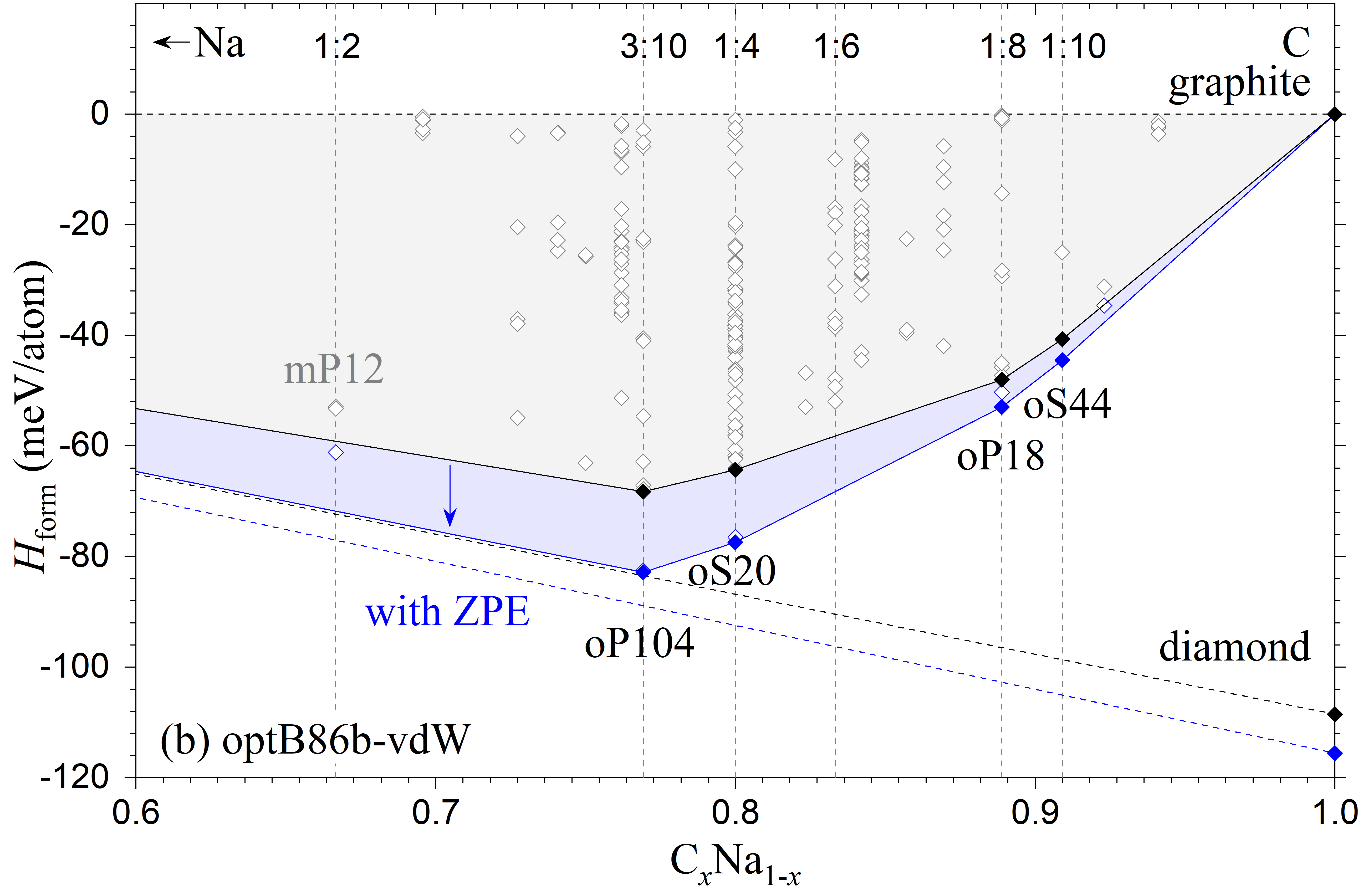}} 
    \caption{Stability of Na-C phases at 10 GPa calculated with (a) the optB88-vdW functional or (b) the optB86b-vdW functional. The global (local) convex hulls are denoted with solid (dashed) lines. The formation enthalpies are shown with and without zero-point energy (ZPE) in blue and gray, respectively. }
    \label{fig:hulls}
\end{figure*}

To investigate e-ph interactions and superconducting properties of oS20-NaC$_4$ further, we employed the EPW code~\cite{Giustino2007,Ponce2016,Margine2013,Lee2023}. The electronic wavefunctions required for the Wannier interpolation~\cite{Marzari2012,Pizzi2020,Marazzo2023} were obtained on a uniform $\Gamma$-centered $8 \times 8 \times 8$ $k$-grid. We used ten atom-centered orbitals to describe the electronic structure of oS20-NaC$_4$, with one $s$ orbital for each Na atom and one $p_z$ orbital for each C atom. The anisotropic Migdal-Eliashberg equations were solved on fine uniform $80 \times 80 \times 80$ $k$- and $40 \times 40 \times40$ $q$-point grids, with an energy window of $\pm 0.2$~eV around the Fermi level and a Matsubara frequency cutoff of 1.0~eV. When solving the isotropic ME equations, the Dirac deltas of electrons and phonons were replaced by Gaussians of width 50~meV and 0.3~meV, respectively.

\section{\label{sec:results}Results and Discussions}

\subsection{Structure and stability of Na-C compounds} 

The global evolutionary searches by Hao {\it et al.}~\cite{Hao2023} provide an important baseline for viable morphologies in the Na-C binary up to 10 GPa. Namely, only graphite configurations intercalated with either one or two layers of Na appear to compete for thermodynamic stability across the composition range. We reproduce the negative formation enthalpy values evaluated with the optB88-vdW functional for the reported oS20-NaC$_4$ ($-17.3$ meV/atom) and mP12-NaC$_2$ ($-15.9$ meV/atom) phases relative to bcc-Na and diamond at 10 GPa. At the same time, we observe a high sensitivity of the results to phonon contributions and DFT approximations due to the starkly different bonding types displayed by the binary and reference phases. The inclusion of the zero-point energy (ZPE) alone shifts the values to $-36.2$~meV/atom and $-28.4$~meV/atom, respectively, and the vibrational entropy further stabilizes the intercalated compounds (see Fig.~\ref{fig:hulls}). The beneficial effect of temperature may be consequential because the formation enthalpies calculated in the optB86b-vdW approximation are positive, at 28.2~meV/atom and 23.7~meV/atom and the formation Gibbs free energies turn negative only at about 600~K. Considering the wide range of enthalpy values of graphite relative to diamond, 59, 88, and 116 meV/atom produced by the optB88-vdW, r$^2$SCAN+rVV10, and optB86b-vdW functionals, respectively, it is evident that the systematic errors are too large to make a definitive conclusion about the global stability of the Na-C intercalated phases.

Fortunately, the diamond reference is likely irrelevant if graphite is used as the starting material in the synthesis of the proposed binary compounds and the sample is not heated to high temperatures required to induce the $sp^2$ to $sp^3$ rebonding. According to the comprehensive review of extensive experimental work on carbon~\cite{Sundqvist2021}, first signs of a structural phase transition in graphite under standard quasi-hydrostatic catalyst-free conditions at room temperature appear at 14$-$17 GPa. Measurements of electrical, optical, X-ray diffraction, and Raman response along with {\it ab initio} simulations indicate that the resulting cold compressed graphite features an $sp^3$-bonded morphology that reverts to the original state upon pressure release. Full irreversible transition to cubic or hexagonal diamond requires heating the sample to high temperatures, typically above 2,000~K at 15~GPa, or applying shear stress. The findings highlight the defining role of kinetics in the evolution of the covalently bonded framework under such moderate pressures. The increase in the interlayer spacing and the transfer of the negative charge to the C layers caused by Na intercalation may affect the kinetic barriers but are not expected to appreciably facilitate the rebonding in the targeted $(T, P)$ range. Therefore, we find it more appropriate to define the synthesizability of the GICs with respect to bcc-Na and graphite rather than diamond not just up to 5 GPa where graphite is thermodynamically stable according to the optB88-vdW results at 0 K~\cite{Hao2023} (up to 2~GPa in our optB86b-vdW calculations) but up to at least 10~GPa. Figure~2 shows that the two proposed phases are indeed thermodynamically favorable with respect to the C layered polymorph, and the relative enthalpies agree well in the optB86b-vdW and optB88-vdW approximations due to the similarity of the elemental and binary morphologies.

Our screening of larger-sized intercalated structures offers further insight into what phases may form under pressure in this binary system. Starting with 14 different supercells of the MgB$_2$ or mP12-NaC$_2$ prototypes, we generated $\sim$ 750 unique Na$_x$C$_2$ configurations and fully optimized them at 10~GPa with both optB88-vdW and optB86b-vdW functionals to construct the convex hulls in Fig.~\ref{fig:hulls} (the common ground states in the two approximations were also evaluated with the r$^2$SCAN+rVV10 functional). The considered phases had hexagonal (7\%), orthorhombic (48\%), monoclinic (38\%), or trigonal (7\%) unit cells with up to 52 atoms at 35 compositions below 1:2. 

At the 1:4 stoichiometry corresponding to the previously proposed oS20 ground state, we find an alternative mP20 structure that is nearly degenerate in Gibbs free energy with oS20 in both optB86b-vdW and optB88-vdW calculations (see Figs.~S1 and S2~\cite{SM}). The two phases have fairly uniform metal decorations but the monoclinic variant features a pronounced interlayer shift that positions Na atoms near hexagon centers in one of the two sandwiching C layers (Figs.~\ref{fig:struct}(c) and~\ref{fig:struct}(d)). The structural frustration with respect to sheer in these and other examined Na-C phases results in soft phonon modes that are difficult to converge. 

\begin{figure}[t]
    \centering \includegraphics[width=8.65cm]{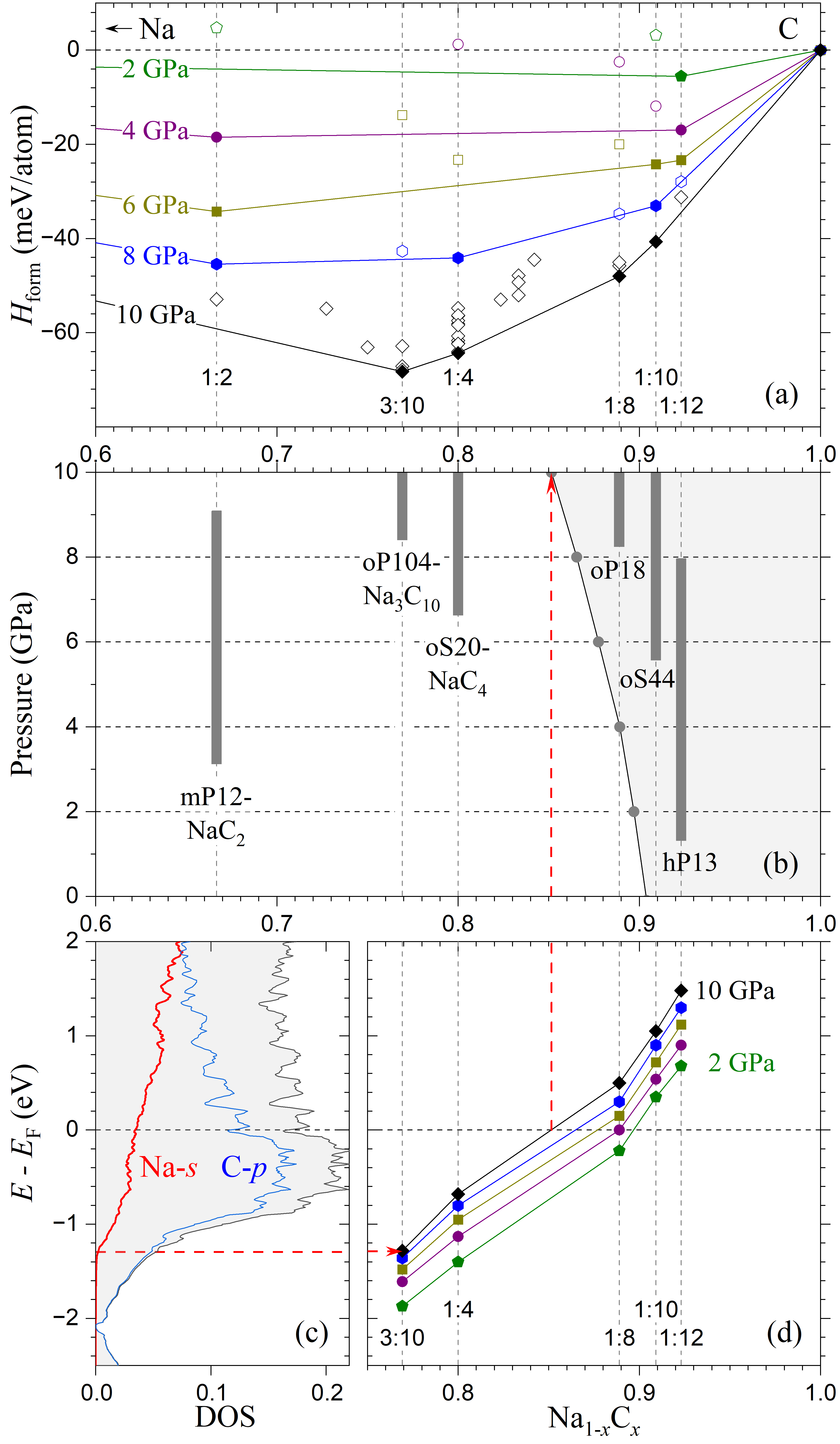}
    \caption{Stability and electronic properties of Na$_{1-x}$C$_x$ compounds calculated with the optB86b-vdW functional. (a) Convex hulls with respect to bcc-Na and graphite at 2-10 GPa. (b) Stability ranges of the stable Na-C phases. The shaded area marks the $(x,P)$ phase space in which Na-C materials are not expected to have electron-doped Na-$s$ states important for superconductivity. (c) The density of states in oP104-Na$_3$C$_{10}$. The dashed red line shows the bottom of the Na-$s$ states. (d) Position of the Na-$s$ band edge for the intercalated Na-C compounds at 2$-$10 GPa. The dashed red line marks an approximate composition at which the Na-$s$ band edge is expected to align with $E_{\rm F}$ at 10 GPa.}
    \label{fig:map}
\end{figure}

\begin{figure}[t]
    \centering \includegraphics[width=8.64cm]{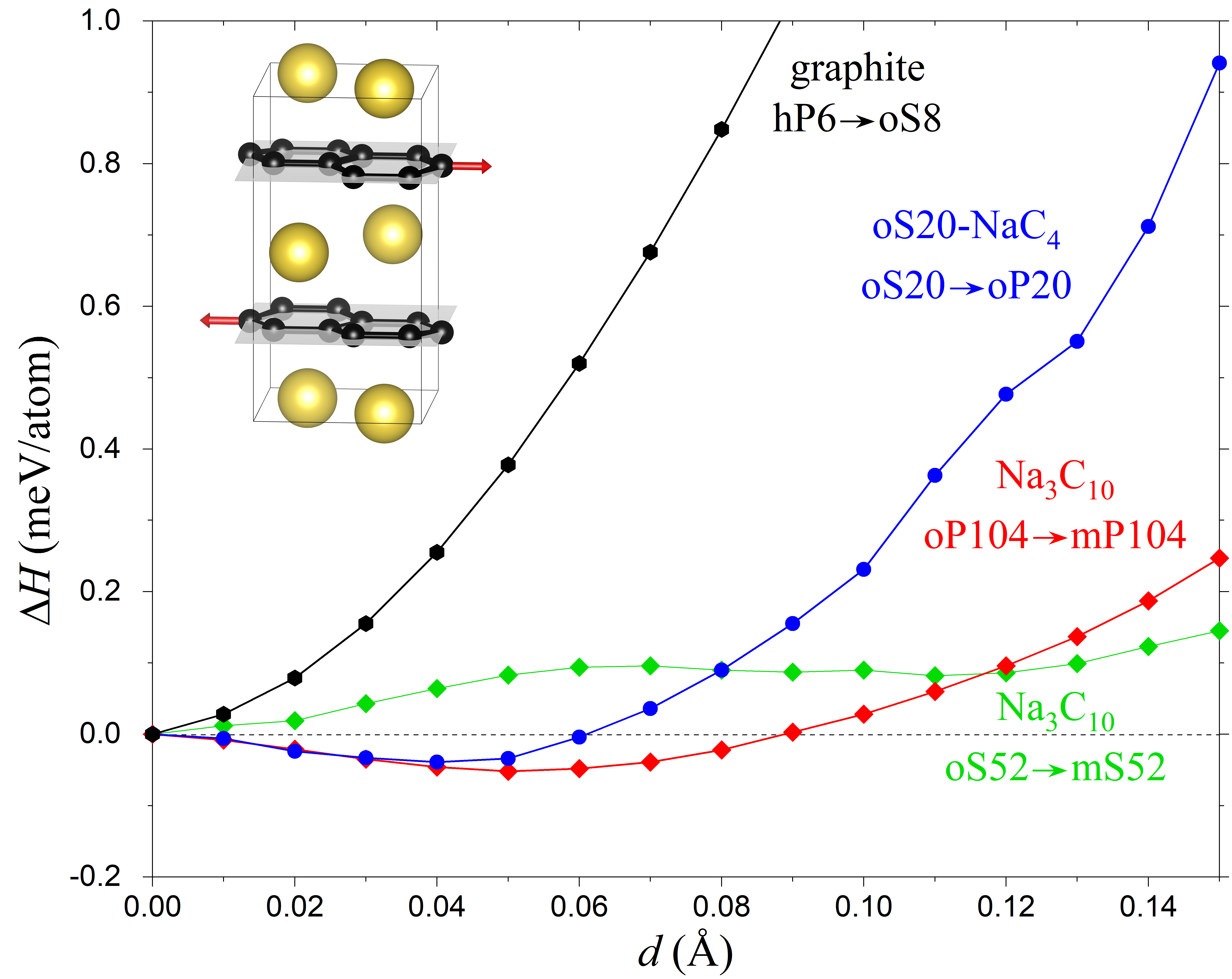}
    \caption{Enthalpy change at 10~GPa along an optical phonon mode eigenvector that shears carbon layers. For each displacement value $d$, the neighboring C layers are shifted along C-C bonds in the opposite directions, with one C atom per layer fixed and all other structural parameters fully optimized. The corresponding changes in symmetry are $P6_3/mmc$ to $Cmcm$ (graphite), $Cmcm$ to $Pca2_1$ (oS20-NaC$_4$), $P2_12_12_1$ to $P2_1$ (oP104-Na$_3$C$_{10}$), and $Cmcm$ to $P2$ (oS52-Na$_3$C$_{10}$).}
    \label{fig:enthalpy}
\end{figure}

The dominant compound at 10 GPa actually occurs at an unusual 3:10 composition in all considered vdW approximations (Figs.~\ref{fig:hulls} and S3~\cite{SM}). The best oS52 structure with the $Cmcm$ symmetry in the generated pool had multiple imaginary modes across the Brillouin zone (see Fig.~S5~\cite{SM}). General strategies to deal with dynamically unstable structures include (i) checking convergence criteria to ensure that the structure indeed has imaginary frequencies~\cite{Pallikara2022,ak33,Lin2022}; (ii) following eigenvector(s) in the search for nearby local minima~\cite{Pallikara2022,ak24,Heil2017,Ying2018}; and (iii) considering anharmonic corrections to stabilize the structure at finite temperatures~\cite{Pallikara2022,Monacelli2021,Zacharias2023,Lucrezi2024b,Jiang2024}. We checked that several phonon modes remain imaginary in our VASP calculations upon further structural relaxation (with residual forces below 0.01~eV/\AA), increase of the $k$-point density (to $6\times 6 \times 6$ in the $2\times 2 \times 1$ 104-atom supercell), or change of the displacement value (0.04~\AA\, or 0.10~\AA). We proceeded with the generation of distorted derivatives by displacing atoms along selected eigenvectors in conventional (super)cells and constructed alternative phases, oS52 ($C222_1$) and oP104 ($P2_12_12_1$), with essentially identical enthalpies (within 0.1~meV/atom). In contrast to our previous successful identification of several local minima in CaB$_6$ by considering all possible combinations of imaginary eigenvectors at two high-symmetry points~\cite{ak24}, the two derived Na$_3$C$_{10}$ polymorphs also displayed several branches with imaginary frequencies (Figs.~S6 and S7~\cite{SM}). To illustrate the particular challenge of finding a dynamically stable structure in this case, we plot the enthalpy profiles along a $\Gamma$-point soft-mode eigenvector representing the collective shear of the C layers in several phases (Fig.~\ref{fig:enthalpy}). The reason behind the much smaller enthalpy variation in the Na-C phases compared to that in graphite is the near 50\% increase in the interlayer spacing that exponentially reduces the corrugation arising from the overlap of the C-$p_z$ orbitals~\cite{ak06}. The resulting barriers in these GICs appear to be defined by the relative Na and C distributions in the neighboring layers. Since the Na atoms are not in perfect registry with the honeycomb lattice in either oS20-NaC$_4$ or oP104-Na$_3$C$_{10}$ ({\it e.g.}, only a third of them reside above and below the C hexagon centers in the latter phase), the lack of short-period commensuration leads to the particularly smooth potential energy surfaces~\cite{ak03,ak05}, and one can think of the materials as bearings with incompressible sheets rolling between large metal ion spheres. These observations help explain the slight differences between frozen phonon and linear response results (Figs.~S7 and S8(b)~\cite{SM}), as the choice of the displacement value can significantly affect the evaluation of force constants. Therefore, it is unlikely that either the creation of larger supercells or the inclusion of anharmonic effects would provide a definitive answer as to whether the layered Na-C phases can be represented with well-defined dynamically stable structures~\cite{Lucrezi2024b}. Domain formation or stacking disorder could be inherent features in these binary compounds, as was predicted in the related LiB material~\cite{ak08,ak09} and supported with XRD measurements~\cite{ak30}. For these reasons, we use the slightly favored oS20 and oP104 unit cells as representative models of the frustrated NaC$_4$ and Na$_3$C$_{10}$ compounds, respectively, in the thermodynamic stability analysis but check the robustness of our superconductivity predictions by examining the additional closely related polymorphs.

The oP104-Na$_3$C$_{10}$ destabilizes the proposed double-layer mP12-NaC$_2$ phase at 10 GPa but leaves room for possible GICs at the low-Na composition end, such as NaC$_8$ and NaC$_{10}$. Figures~\ref{fig:struct}(e) and~\ref{fig:struct}(f) illustrate that the most favored configurations feature nearly flat C layers and Na intercalating every other gallery, known as stage-two GICs. The lower Na concentration allows the metal sublattice to find a better balance between even distribution and registry with the honeycomb lattice. Consequently, oP18-NaC$_8$ and oS44-NaC$_{10}$ are found to be dynamically stable (see Fig.~\ref{fig:all-ph-a2f}(f) and (g)).

The resulting local convex hull (relative to graphite) at 0 K and 10 GPa is defined by four phases at the 3:10, 1:4, 1:8, and 1:10 compositions in all optB86b-vdW, optB88-vdW, and r$^2$SCAN+rVV10 approximations (Figs.~\ref{fig:hulls} and~S3~\cite{SM}). The set of ground states under 10~GPa compression remains unchanged at elevated temperatures once the vibrational entropy is included (Figs.~S1 and S2~\cite{SM}), while transition pressures evaluated for select phases at 600~K instead of zero~K shift by $-0.8$~GPa for oP104-Na$_3$C$_{10}$, $-1.5$~GPa for oS20-NaC$_4$, and 0.7~GPa for oP18-NaC$_8$. Interestingly, oP104-Na$_3$C$_{10}$ is the only intercalated phase that is globally stable at room temperature in both DFT treatments (Fig.~\ref{fig:hulls}). To probe what GICs could be stable in the $1-10$~GPa range, we reoptimized relevant phases at specific pressures, examined their relative enthalpies (Fig.~S4~\cite{SM}), and constructed corresponding local convex hulls at 0~K [Fig.~\ref{fig:map}(a)]. The first material to stabilize under compression is a stage-two hP13-NaC$_{12}$ GIC derived from the $\sqrt{3}\times\sqrt{3}\times2$ supercell of MgB$_2$. mP12-NaC$_2$ may indeed form just above 3~GPa (a related LiB compound with double metal layers has been synthesized via cold compression above 23~GPa~\cite{ak30}). The four proposed materials between the 3:10 and 1:10 ratios require higher pressures to be viable [see Fig.~\ref{fig:map}(b)]. Our comparison of the optB86b-vdW and optB88-vdW results in Fig.~S4~\cite{SM} shows that the phase transition pressure estimates in the two approximations agree to within 1~GPa.

\begin{figure*}[!htb]
    \centering
    \includegraphics[scale=0.78]{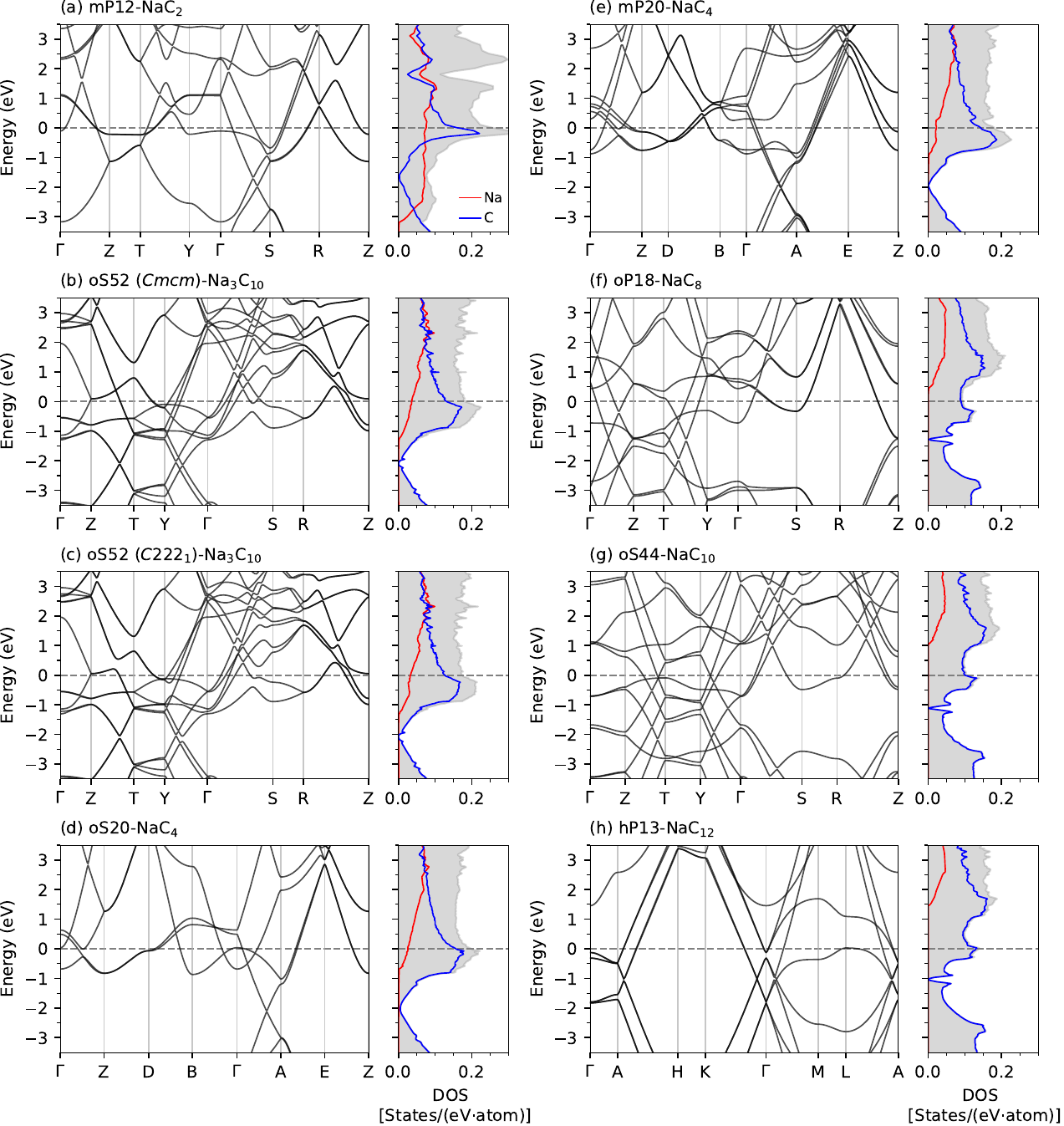}
    \caption{Electronic band structures and density of states (DOS) calculated with Quantum {\small ESPRESSO} at 10~GPa for (a) mP12-NaC$_{2}$, (b) oS52 ($Cmcm$)-Na$_3$C$_{10}$, (c) oS52 ($C222_1$)-Na$_3$C$_{10}$, (d) oS20-NaC$_{4}$, (e) mP20-NaC$_{4}$, (f) oP18-NaC$_8$, (g) oS44-NaC$_{10}$, and (h) hP13-NaC$_{12}$.}
    \label{fig:all-ele-band}
\end{figure*}

\begin{figure*}[!htb]
    \centering
    \includegraphics[scale=0.76]{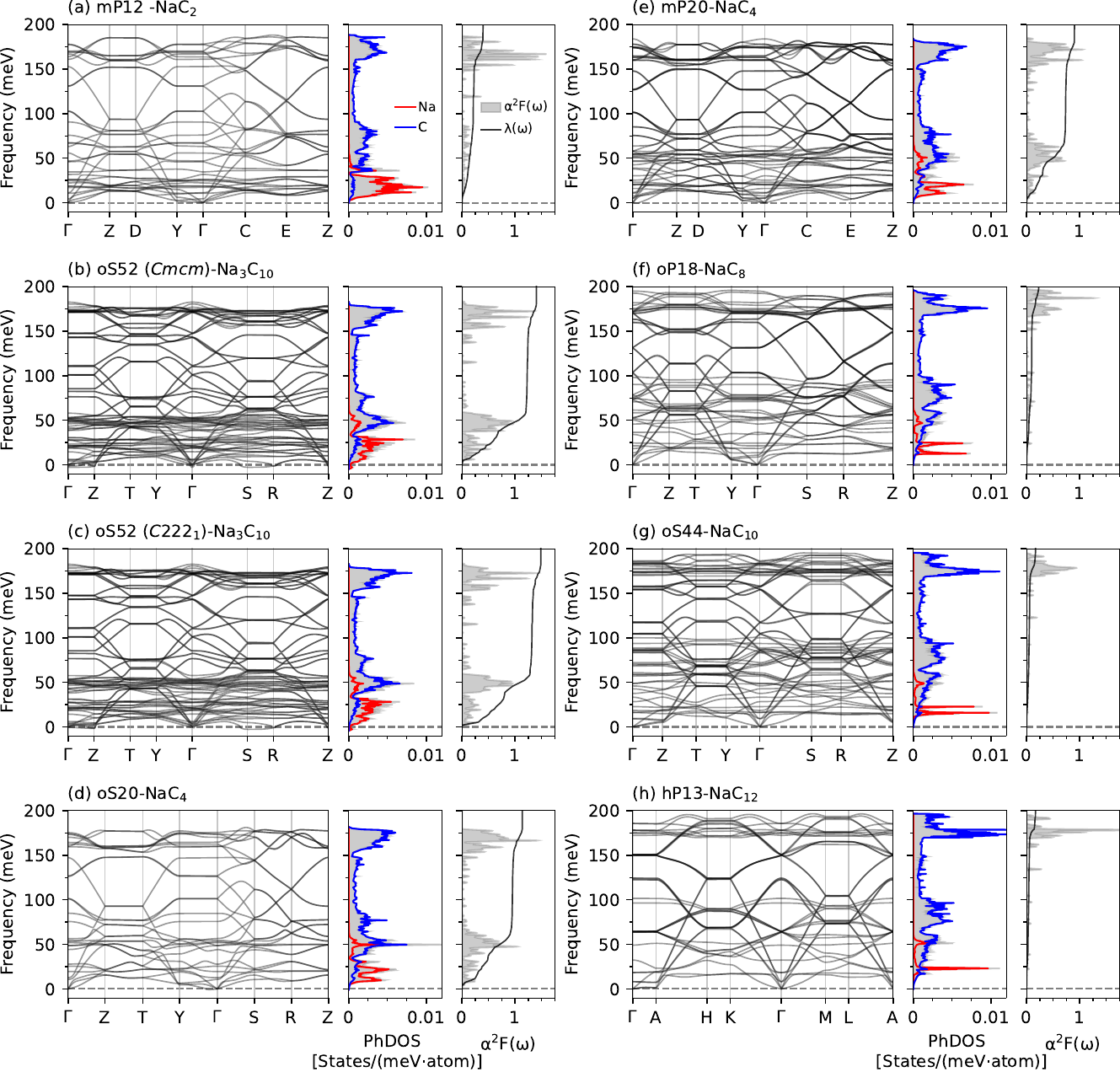}
    \caption{Phonon dispersion, phonon density of states (PhDOS), Eliashberg spectral function $\alpha^2 F(\omega)$, and integrated electron-phonon coupling strength $\lambda^{\rm QE}$ calculated with Quantum {\small ESPRESSO} at 10~GPa for (a) mP12-NaC$_{2}$, (b) oS52 ($Cmcm$)-Na$_3$C$_{10}$, (c) oS52 ($C222_1$)-Na$_3$C$_{10}$, (d) oS20-NaC$_{4}$, (e) mP20-NaC$_{4}$, (f) oP18-NaC$_8$, (g) oS44-NaC$_{10}$, and (h) hP13-NaC$_{12}$. The $\alpha^2F(\omega)$ and $\lambda^{\rm QE}$ for Na$_3$C$_{10}$ structures are estimated by neglecting the imaginary phonon frequencies.}
    \label{fig:all-ph-a2f}
\end{figure*}

\begin{table*}[!htb]
    \caption{Properties of Na$_{1-x}$C$_x$ phases: the average interlayer distance $d_{\rm inter}$, the density of states at the Fermi level $N(E_{\rm F})$, the nearly-free energy (NFE) band occupation, the logarithmic average phonon frequency $\omega_{\log}$, the total electron-phonon coupling strength $\lambda^{\rm QE}$, and the superconducting critical temperature $T_{\rm c}^{\rm QE}$ with a Coulomb potential $\mu^*=0.1$. Note that $\lambda^{\rm QE}$ and $T_{\rm c}^{\rm QE}$ for oP104-Na$_3$C$_{10}$ were obtained using phonon calculations for a single $q$-point ($\Gamma$).} 
\label{tab:supercond-QE}
\setlength\tabcolsep{0pt} % let LaTeX compute intercolumn whitespace
\smallskip 
\begin{tabular*}{\textwidth}{@{\extracolsep{\fill}}l c c c c c  c c c c}
\hline\hline \noalign{\vskip 1mm}
 Phase  & Space  & Pearson  & Pressure & $d_{\rm inter}$ & $N(E_{\rm F})$ & NFE  & $\omega_{\log}$ & $\lambda^{\rm QE}$ & $T_{\rm c}^{\rm QE}$  \\
   composition    & group  & symbol & (GPa) &  (\AA)  &  (states/eV$\cdot$atom) &   occupation   &    (meV)  &      & (K) \\ \noalign{\vskip 1mm}
\hline                      
        NaC$_{2}$      & $P2_1m$      & mP12  & 10 & 6.81 & 0.261 & Yes & 64.1  & 0.40 & 3.0 \\ 
        Na$_3$C$_{10}$ & $Cmcm$       & oS52  & 10 & 4.22 & 0.191 & Yes & 30.4  & 1.40 & 37.5\\ 
        Na$_3$C$_{10}$ & $C222_1$     & oS52  & 10 & 4.20 & 0.181 & Yes & 22.5  & 1.50 & 29.6 \\ 
        Na$_3$C$_{10}$ & $P2_12_12_1$ & oP104 & 10 & 4.26 & 0.176 & Yes & 30.8  & 1.43 & 39.1 \\ 
        NaC$_{4}$      & $Cmcm$       & oS20  & 10 & 4.23 & 0.199 & Yes & 34.6  & 1.13 & 33.5 \\
        NaC$_{4}$      & $P2_1/m$     & mP20  & 10 & 4.17 & 0.161 & Yes & 42.3  & 0.89 & 28.5 \\ 
        NaC$_8$        & $Pmma$       & oP18  & 10 & 4.21 & 0.066 & No  & 97.8  & 0.23 & - \\ 
        NaC$_{10}$     & $Cmcm$       & oS44  & 10 & 4.17 & 0.085 & No  & 108.8 & 0.17 & - \\ 
        NaC$_{12}$     & $P6/mmm$     & hP13  & 5  & 4.10 & 0.153 & No  & 114.5 & 0.18 & - \\ 
        NaC$_{12}$     & $P6/mmm$     & hP13  & 10 & 4.10 & 0.141 & No  & 120.9 & 0.17 & - \\ \noalign{\vskip 1mm}  
\hline\hline
\end{tabular*}
\end{table*}

\subsection{Electronic, vibrational, and superconducting properties of Na-C phases}

To screen the Na-C binary system for possible high-$T_{\rm c}$ superconductivity, we investigated the electronic, vibrational, and superconducting properties in the proposed set of intercalated phases, mP12--NaC$_2$, oS52 (oP104)-Na$_3$C$_{10}$, oS20 (mP20)-NaC$_4$, oP18-NaC$_8$, oS44-NaC$_{10}$, and hP13-NaC$_{12}$, at pressures up to 10 GPa. First, we examined the response of the intercalant Na-$s$ states to composition and pressure to determine conditions under which the band crosses the Fermi level ($E_{\rm F}$). The presence of the partially occupied nearly free electron (NFE) states has been linked to superconductivity in several layered compounds ({\it e.g.}, CaC$_6$ and YbC$_6$ GICs~\cite{Calandra2005, Csanyi2005, Mazin2005}, bilayer C$_6$CaC$_6$~\cite{Mazin2010,Margine2016}, monolayer LiC$_6$~\cite{Profeta2012,Zheng2016}, and LiB~\cite{Calandra2007, Kafle2022}). The strong interaction of this band with the surrounding lattice has been rationalized as a response to both quantum confinement and electrostatic effects~\cite{Profeta2012, Boeri2007, Margine2006}.

Our electronic band structure and density of states (DOS) calculations show that only compounds at the 1:2, 3:10, and 10:4 compositions have partially occupied Na-$s$ states in the considered pressure range [see Figs.~\ref{fig:map}(c), (d) and \ref{fig:all-ele-band}]. Among them, mP12-NaC$_2$ stands out as the only one with a 2D shape of the Na DOS around the Fermi level and with very little dependence of the band edge position on pressure (about $-3.0$~eV at 10 GPa). In contrast, just as in LiB~\cite{Calandra2007}, the Na DOS in the Na$_3$C$_{10}$ and NaC$_4$ phases has a 3D profile, and compression from 2~GPa to 10~GPa raises these NFE states by about 0.6~eV. Fortunately, the band edge remains at $-1.3$~eV and $-0.7$~eV below the Fermi level in the two respective Na-C compounds at 10~GPa as shown in Fig.~\ref{fig:map}(d). As in CaC$_6$~\cite{Sanna2007,Calandra2005}, the dominant contribution to the DOS at the Fermi level ($N(E_{\rm F})$) comes from the anti-bonding C-$\pi^*$ states. Another important characteristic is the close resemblance of the overall DOS shape in the oS52 (oP104)-Na$_3$C$_{10}$ and oS20 (mP20)-NaC$_4$ phases at the 3:10 and 1:4 compositions, suggesting that they may have similar superconducting properties.

Decreasing the metal concentration has a strong effect on the position of the NFE band. The data in Figs.~\ref{fig:map}(d) and \ref{fig:all-ele-band} illustrate that while the stage-two oP18-NaC$_8$, oS44-NaC$_{10}$, and hP13-NaC$_{12}$ are still metallic, none of them features the Na-$s$ states at the Fermi level in the corresponding regions of thermodynamic stability. For instance, this band would be partially occupied in oP18-NaC$_8$ below 4~GPa but the phase requires over 8~GPa to break the convex hull (see Fig.~\ref{fig:map}(b)). Since the three Na-poor GICs have similar DOS shapes and can be modeled with a rigid-band approximation, $N(E_{\rm F})$ is defined primarily by the amount of available charge to fill the C-$\pi^*$ states and happens to decrease with the Na content. Figure~\ref{fig:all-ele-band}(f-h) and Table~\ref{tab:supercond-QE} demonstrate that the resulting $N(E_{\rm F})$ values in two of the three phases are significantly lower than those in the considered stage-one GICs.

To construct an approximate $x(P)$ boundary between Na$_{1-x}$C$_x$ compounds with and without populated Na-$s$ states, we estimated the critical concentrations $x$ by linear interpolations at every 2 GPa pressure (the dashed red line across panels (b) and (d) of Fig.~\ref{fig:map} illustrates the determination of the point at 10 GPa). The near-linear shift of the critical compositions with pressure found for the phases with different symmetries suggests little sensitivity of the Na-$s$ energies to the particular decoration of the metal sites as long as the arrangements within galleries are fairly uniform (see Fig.~\ref{fig:struct}). The resulting shaded area in Fig.~\ref{fig:map}(b), extrapolated down to the ambient pressure, marks the $x(P)$ region in which Na-C phases are not expected to have superconductivity arising primarily from the strong coupling to the intercalant states.

To check the validity of this observation, we looked at the vibrational and superconducting properties for this set of materials with the QE package. Similar to CaC$_6$ and several other GICs~\cite{Calandra2005,Calandra2007b,Sanna2012}, 
the phonon dispersion can be divided into three regions (see Figs.~\ref{fig:all-ph-a2f}, S8 and S11~\cite{SM}): a low-frequency region composed of mostly in-plane Na$_{xy}$ vibrations mixed with out-of-plane C$_z$ vibrations ($< 30$~meV), an intermediate frequency region dominated by out-of-plane C$_z$ and Na$_z$ vibrations ($30-100$~meV), and a high-frequency region of in-plane C$_{xy}$ vibrations ($> 100$~meV). The superconducting $T_{\rm c}$ was estimated using the Allen-Dynes modified McMillan formula~\cite{McMillan1968,Allen1975} with a Coulomb potential $\mu^*=0.1$. We denote the e-ph coupling strength and the critical temperature obtained with QE as $\lambda^{\rm QE}$ and $T_{\rm c}^{\rm QE}$, respectively, to differentiate them from the EPW values discussed in the next section. Table~\ref{tab:supercond-QE} summarizes $\lambda^{\rm QE}$ and $T_{\rm c}^{\rm QE}$ for all the Na$_{1-x}$C$_x$ compounds.

Unsurprisingly, the three phases in the gray region of Fig.~\ref{fig:map}(b) probed within their respective stability regions oP18-NaC$_8$ at 10~GPa, oS44-NaC$_{10}$ at 10~GPa, and hP13-NaC$_{12}$ at 5~GPa, were found to have low $\lambda^{\rm QE}$ of 0.23, 0.17, and 0.18, respectively, with no measurable $T_{\rm c}^{\rm QE}$ values. Upon examination of the Eliashberg spectral function $\alpha^2F(\omega)$ and the cumulative e-ph coupling strength $\lambda(\omega)$ plotted in Fig.~\ref{fig:all-ph-a2f}, we conclude that the underlying factor determining the reduced $\lambda^{\rm QE}$ values in these cases is the absence of e-ph coupling for the Na and out-of-plane C$_z$ phonon modes from the low- and intermediate-frequency region below 100~meV.

\begin{figure*}[!t]
    \centering
    \includegraphics[scale=0.94]{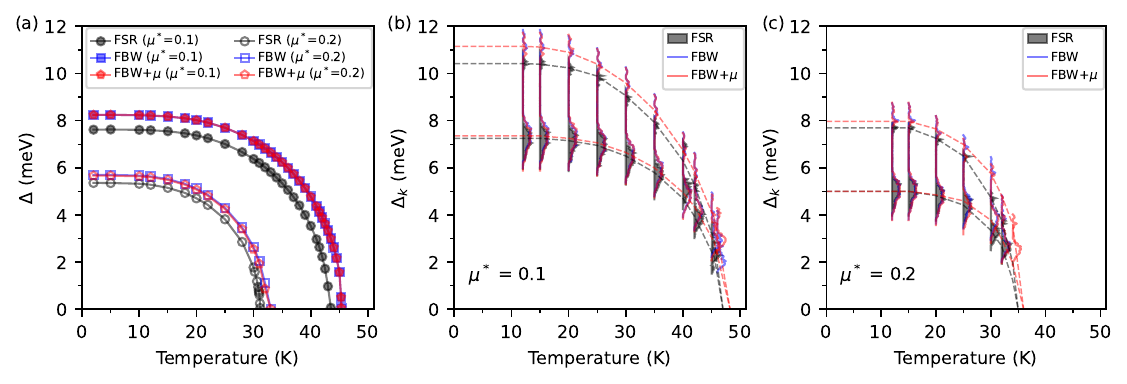}
    \caption{Superconducting properties of oS20-NaC$_4$ at 10~GPa obtained with the FSR (gray), FBW (blue) and FBW+$\mu$ (red) approaches for $\mu^* = 0.1$ and 0.2. (a) Isotropic superconducting gap $\Delta$ as a function of temperature. (b) and (c) Energy distribution of the anisotropic superconducting gap $\Delta_{\bm k}$ as a function of temperature. The black dashed lines in (b) and (c) represent a guide for the eye highlighting the two superconducting gap distributions.}
    \label{fig:gap-aniso}
\end{figure*}

For oS20-NaC$_4$ and mP12-NaC$_2$ at 10 GPa, our calculations produced $\lambda^{\rm QE}$ values of 1.13 and 0.40, with corresponding $T_{\rm c}^{\rm QE}$ values of 33.5~K and 3.0~K, consistent with Ref.~\cite{Hao2023}. We find that the low- and intermediate-frequency phonons play a key role in achieving a high e-ph coupling in oS20-NaC$_4$, and together they account for approximately 84\% of the total $\lambda^{\rm QE} = 1.13$. Moreover, the relative contributions of these two sets of modes are very similar, with the low and intermediate regions making up 40\% and 44\% of $\lambda$, respectively. Notably, the high-frequency in-plane C$_{xy}$ modes exhibit a significantly smaller contribution of only about 16\%. This behavior is similar to other GICs ({\it e.g.}, CaC$_6$, BaC$_6$, SrC$_6$~\cite{Calandra2005,Calandra2007b,Sanna2007,Margine2016}) and Ca-intercalated bilayer graphene~\cite{Margine2016}, where the largest contribution to the e-ph coupling strength has been associated with the modes below 100~meV. While the coupling to these critical modes is present in mP12-NaC$_2$, their contribution is reduced by a factor of four compared to that in oS20-NaC$_4$, which leads to a drastic reduction in the e-ph coupling strength and the critical temperature. For the competing mP20-NaC$_4$ structure at the 1:4 composition, we find $\lambda^{\rm QE}=0.89$ and $T_{\rm c}^{\rm QE}=28.5$~K. Comparison of Figs.~\ref{fig:all-ph-a2f}(d) and \ref{fig:all-ph-a2f}(e) reveals that the in-plane C$_{xy}$ vibrations in the high-frequency region above 100~meV have a nearly identical contribution to $\lambda^{\rm QE}$ in the two NaC$_4$ phases, and the difference in their total coupling strengths can be again attributed to the low- and intermediate-frequency phonons.

Finally, we investigated the superconducting properties of the three Na$_3$C$_{10}$ structures. Since at the harmonic level the phonon dispersions exhibit two slightly imaginary modes with frequencies around $-2.9$~meV, the e-ph properties were computed by integrating only over the real frequency modes (see Figs.~\ref{fig:all-ph-a2f}(b), \ref{fig:all-ph-a2f}(c) and S8(b)~\cite{SM}), similar to what was done in recent studies~\cite{Ferreira2023,Tsuppayakorn2023} (note that the lower integration limit in QE is set to 2.5 meV and such modes would be ignored even if their frequencies were real but below this threshold). Importantly, the phonon DOS and the Eliashberg spectral functions results plotted in Fig.~\ref{fig:all-ph-a2f} show that the soft phonon modes below 10~meV corresponding to C layer shifts have insignificant contribution to the total coupling even in the dynamically stable NaC$_4$ phases and that the dominant parts of 30$-$50\% and 40$-$50\% in all considered 1:4 and 3:10 GICs come from the Na-based modes in the 0$-$30~meV range and C-based modes in the 30$-$100~meV range, respectively. The relatively high $\lambda^{\rm QE}$ ($T_{\rm c}^{\rm QE}$) values of 1.40 (37.5~K), 1.50 (29.6~K), and 1.43 (39.1~K) obtained at 10~GPa for the oS52 ($Cmcm$), oS52 ($C222_1$), and oP104 structural models indicate that Na$_3$C$_{10}$ should be as good a superconductor as NaC$_4$. Hence, we believe that the conclusions drawn from our following in-depth analysis of the superconducting properties in oS20-NaC$_4$ with the anisotropic Migdal-Eliashberg formalism should be transferable to the neighboring Na$_3$C$_{10}$ compound.

\subsection{Superconducting properties of \texorpdfstring{NaC$_4$}{nac4} with the anisotropic Migdal-Eliashberg formalism}

With the oS20-NaC$_4$ phase determined to stabilize above 6.5~GPa, we performed a more detailed examination of its properties at 10~GPa. More insight into the electronic structure can be obtained by analyzing the dispersion of oS20-NaC$_4$ with Na atoms removed (empty graphite-like C$_4$ structure, solid black lines) and the one with C atoms removed (empty Na metal structure, red dashed lines) depicted in Fig.~S9(c)~\cite{SM}. Comparing Figs.~S9(a) and~S9(c)~\cite{SM}, we can interpret the band structure of oS20-NaC$_4$ as a superposition of the C$_4$ and Na bands. To facilitate this comparison, the C and Na bands in Fig.~S9(c) are shifted downward and upward by $\sim$2~eV, respectively, to match the Fermi level in oS20-NaC$_4$. Effectively, the relative shift is equivalent to a charge transfer from the Na atoms to the graphene layers. This can be clearly visualized from the charge density difference plot in Fig.~S9(d), where the red and green lobes represent the charge accumulation and depletion regions around the C and Na atoms, respectively. Integrating the atomic-projected DOS, we extract the L\"{o}wdin charge for each atom. In oS20-NaC$_4$, we find a charge transfer of 0.21~electrons/C atom, whereas in CaC$_6$ each C atom only gains 0.11~electrons~\cite{Calandra2005}.

Figure~S10~\cite{SM} illustrates the total and band-decomposed Fermi surface (FS). Comparison with the band structure in Fig.~S9(a) reveals that the three bands crossing the Fermi level give rise to a multi-sheet FS structure. The FS sheets corresponding to band 1 (shown in blue) and band 2 (shown in green) exhibit 3D nested shapes and originate from the C-$p_z$ states, as already pointed out in Ref.~\cite{Hao2023}. These nested regions surround two bean-shaped FS sheets (shown in red) centered around $\Gamma$ of predominant Na-$s$ character (band 3). 

\begin{table*}[!hbt]
\small
\footnotesize
    \centering
    \begin{tabular}{c c c c c c c c c c c c}
        \hline \hline
         $N(E_{\rm F})$ & \hspace*{2.0mm} $\omega_{\rm log}$ \hspace*{2.0mm} 
         & \hspace*{2.0mm} $\lambda$ \hspace*{2.0mm} 
         & \hspace*{2.0mm} $\mu^*$ \hspace*{2.0mm} &
         \multicolumn{2}{c}{\hspace*{0.5mm}Semi-empirical $T_{\rm c}$~(K)\hspace*{0.5mm}} &
        \multicolumn{3}{c}{\hspace*{2.0mm} Isotropic ME $T_{\rm c}$~(K) \hspace*{2.0mm}}&
        \multicolumn{3}{c}{\hspace*{2.0mm} Anisotropic ME $T_{\rm c}$~(K) \hspace*{2.0mm}} \\
         (states/(eV$\cdot$atom)) & (meV) & & & \hspace*{0.5mm} AD \hspace*{0.5mm}
         & \hspace*{0.5mm} ML \hspace*{0.5mm} & 
         \hspace*{0.5mm} FSR \hspace*{0.5mm} & 
         \hspace*{0.5mm} FBW \hspace*{0.5mm} & 
         \hspace*{0.5mm} FBW+$\rm \mu$ \hspace*{0.5mm} & 
         \hspace*{0.5mm} FSR \hspace*{0.5mm} & 
         \hspace*{0.5mm} FBW \hspace*{0.5mm} & 
         \hspace*{0.5mm} FBW+$\rm \mu$ \hspace*{0.5mm} \\
        \hline 
        \multirow{2}{*}{0.19} & \multirow{2}{*}{33.9} & 
        \multirow{2}{*}{1.19} & 0.1 & 34.9 & 43.3 & 43.5 & 45.5 & 45.5 & 47.0 & 48.2 & 48.2\\
        %\hline 
         &  &  & 0.2 & 22.0 & 26.7 & 31.2 & 33.0 & 33.0 & 35.0 & 36.0 & 36.0\\
        \hline \hline
    \end{tabular}
    \caption{Properties of oS20-NaC$_4$ at 10~GPa: the density of states at the Fermi level $N(E_{\rm F})$, the logarithmic average phonon frequency $\omega_{\rm log}$, the electron-phonon coupling strength $\lambda$, the Coulomb parameter $\mu^*$, and the superconducting critical temperature $T_{\rm c}$. The $T_{\rm c}$ values are calculated using different methodologies: the semi-empirical Allen-Dynes (AD) formula, the machine-learned (ML) model, the isotropic and anisotropic Fermi surface restricted (FSR) approximation, and the isotropic and anisotropic full bandwidth implementation without (FBW) and with (FBW+$\mu$) variation of the chemical potential $\mu$.} 
    \label{tab:Tc}
\end{table*}

To investigate the e-ph interaction, we computed the isotropic Eliashberg spectral function $\alpha^2F(\omega)$ and the cumulative e-ph coupling strength $\lambda(\omega)$ with EPW (see Fig.~S11(c)~\cite{SM}). Consistent with our QE results, we find a high e-ph coupling strength $\lambda = 1.19$.  The anisotropy in the e-ph coupling is quantified by evaluating the momentum-resolved $\lambda_{\bm k}$ as defined in Ref.~\cite{Margine2013}. Figure~S12(a)~\cite{SM} clearly displays two peaks in the distribution of $\lambda_{\bm k}$, with a sharp peak centered around 1 and a small peak centered around 3. To further analyze the origin of these peaks, the variation of $\lambda_{\bm k}$ on the Fermi surface is shown in Fig.~S12(b)~\cite{SM}. According to the color map, the first sharp peak originates from the nested FS regions, while the second smaller peak arises from the $\Gamma$-centered FS regions. In particular, by comparing with the band structure and the band decomposed FS plots in Figs.~S9(a) and~S10~\cite{SM}, it becomes evident that the first peak arises from the coupling of the phonon modes with the C-$\pi^*$ states (band 1 and band 2), while the second peak is mostly due to the intercalant Na-$s$ states (band 3) with some contribution from the C-$\pi^*$ states (band 2). 

\begin{figure}[!t]
    \centering
    \includegraphics[scale=0.19]{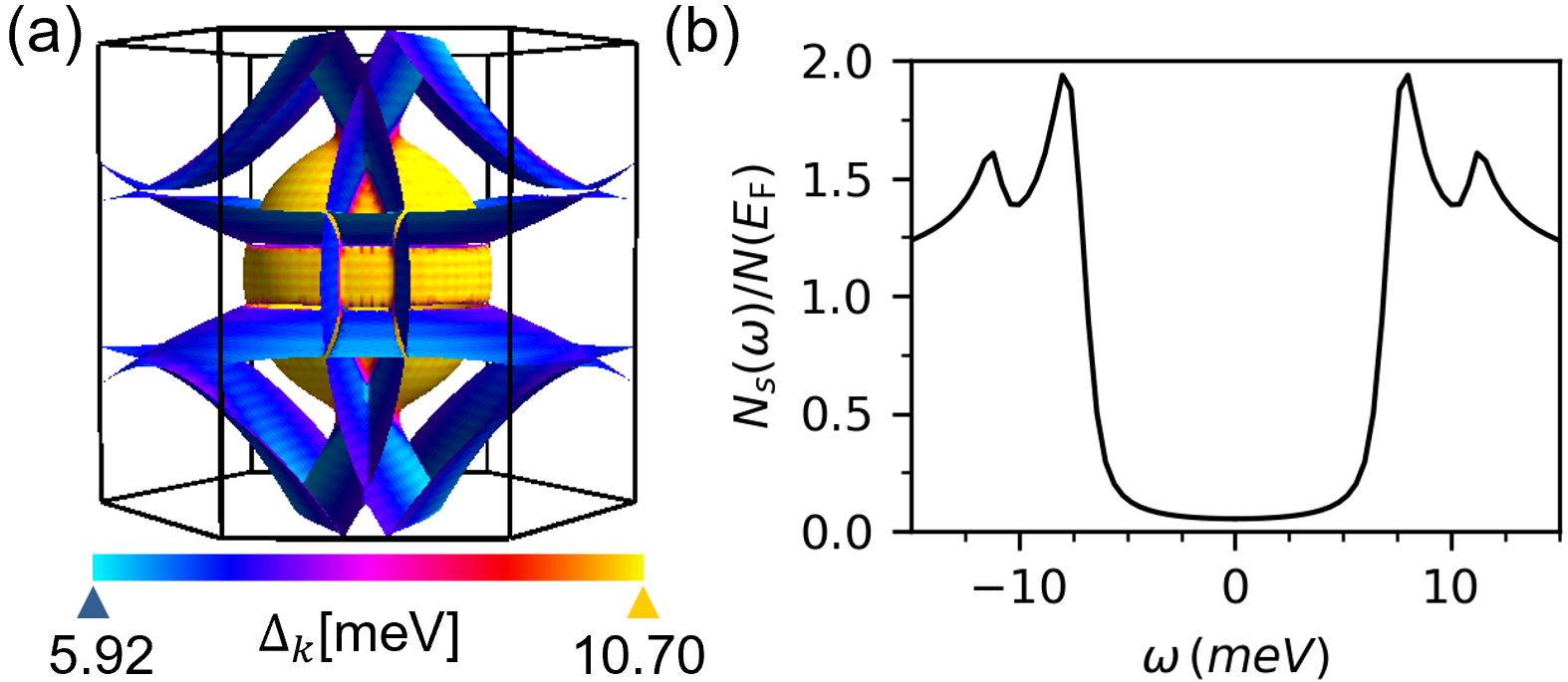}
    \caption{(a) Momentum-resolved superconducting gap $\Delta_{\bm k}$ on the Fermi surface generated with FermiSurfer~\cite{Kawamura2019}, and (b) normalized quasiparticle DOS using the FSR approach with $\mu^*=0.1$ for oS20-NaC$_4$ at 12~K and 10~GPa.} 
    \label{fig:gap-FS}
\end{figure}

In order to determine the superconducting critical temperature, we solved the isotropic and anisotropic Migdal-Eliashberg  equations~\cite{Margine2013,Lee2023}. Since the electronic DOS strongly peaks close to $E_{\rm F}$, we examined the superconducting gap using both the Fermi surface restricted (FSR) and the full bandwidth (FBW) approaches~\cite{Lee2023,Lucrezi2024}. In FSR, the DOS around $E_{\rm F}$ is assumed constant, whereas in FBW the full energy dependence of the DOS is considered, allowing the inclusion of e-ph scattering processes away from $E_{\rm F}$. A recent study has demonstrated the importance of the FBW treatment in the high-pressure H$_3$S and the low-pressure BaSiH$_8$ superconductors, which exhibit distinct DOS features that deviate significantly from the constant-DOS assumption made in the FSR approximation~\cite{Lucrezi2024}. For a detailed description of the FSR and FBW approaches and their implementation in the EPW code, please see Refs.~[\onlinecite{Margine2013,Ponce2016,Lee2023,Lucrezi2024}].

Figure~\ref{fig:gap-aniso}(a) shows the isotropic superconducting gap $\Delta$ as a function of temperature, calculated with the FSR, FBW, and FBW+$\mu$ methods using a Coulomb parameter $\mu^*$ = 0.1 and 0.2. In the FBW+$\mu$ approach, the chemical potential is updated self-consistently as discussed in Ref.~\cite{Lucrezi2024}. Our calculations yield $T_{\rm c}$ values of 43.5 (31.2)~K and 45.5 (33.0)~K, and a zero-temperature superconducting gap of 7.62 (5.36)~meV and 8.24 (5.70)~meV for FSR and FBW with $\mu^*$ = 0.1 (0.2), respectively. Furthermore, the superconducting gap versus temperature curves for FBW and FBW+$\mu$ are basically identical, thus leading to the same $T_{\rm c}$ values. This can be understood as the Fermi level in oS20-NaC$_4$ lies slightly away from the van Hove singularity (vHS) peak, as shown in Fig.S9(b), and therefore the DOS with and without updating $\mu$ are very similar. For comparison, the commonly employed Allen-Dynes (AD) modified McMillan formula~\cite{McMillan1968,Allen1975} and the machine-learned (ML) SISSO model~\cite{Xie2022} give $T_{\rm c}$ values of 34.9 (22.0)~K and 43.3 (26.7)~K for $\mu^*$ = 0.1 (0.2), respectively, with the latter producing results that match very well our isotropic predictions. 

Given the anisotropy found in $\lambda_{\bm{k}}$, we calculated the anisotropic superconducting gap $\Delta_{\bm{k}}$ as a function of temperature shown in Figs.~\ref{fig:gap-aniso}(b) and \ref{fig:gap-aniso}(c). We find two distinct gaps with mean values $\Delta_1 = 7.25$ (5.00)~meV and $\Delta_2 = 10.42$ (7.69)~meV for $\mu^*$ = 0.1 (0.2) in the zero-temperature limit using the FSR approach. For comparison, CaC$_6$ displays a continuum anisotropic gap without separation into distinct gaps~\cite{Margine2016,Sanna2007, Sanna2012}. The $\Delta_1$ gap is highly anisotropic and varies over a wide energy range of 1.75~meV. In contrast, the $\Delta_2$ gap is less pronounced (\textit{i.e.}, fewer energy states contribute to this gap) and has a much narrower distribution of only 0.75~meV. The superconducting $T_{\rm c}$ is estimated to be 47.0 (35.0)~K for $\mu^*$ = 0.1 (0.2), about 8\% (12\%) higher than the isotropic FSR value. Application of the FBW and FBW+$\mu$ treatments only shifts slightly the gap distribution to higher energies while retaining the overall shape at each temperature. The resulting $T_{\rm c}$ using the two FBW approaches is estimated to be 48.2 (36.0)~K with $\mu^*$ = 0.1 (0.2). 
The $T_{\rm c}$ values obtained with different approaches are summarized in Table~\ref{tab:Tc}. 

Figure~\ref{fig:gap-FS}(a) shows the momentum-resolved superconducting gap  $\Delta_{\bm{k}}$ on the FS at 12~K within the FSR approximation using $\mu^*=0.1$. Comparing this with Fig.~S12(b)~\cite{SM}, the regions of the FS with the lower gap $\Delta_1 = 7.35$~meV can be associated with the C-$\pi^*$ states (band 1 and band 2). In contrast, the upper gap $\Delta_2 = 10.60$~meV primarily arises from the Na-$s$ states (band 3) with some contribution from the C-$\pi^*$ states (band 2). Similarly, in CaC$_6$, the smallest and highest values of the smeared multi-gap structure have been associated with the C-$\pi^*$ and Ca-$s, d_{z^2}$ states~\cite{Sanna2007, Sanna2012}. The two-gap structure of oS20-NaC$_4$ is also evident from the corresponding quasiparticle DOS plotted in Fig.~\ref{fig:gap-FS}(b).

\begin{figure}[!t]
    \centering
    \includegraphics[scale=0.95]{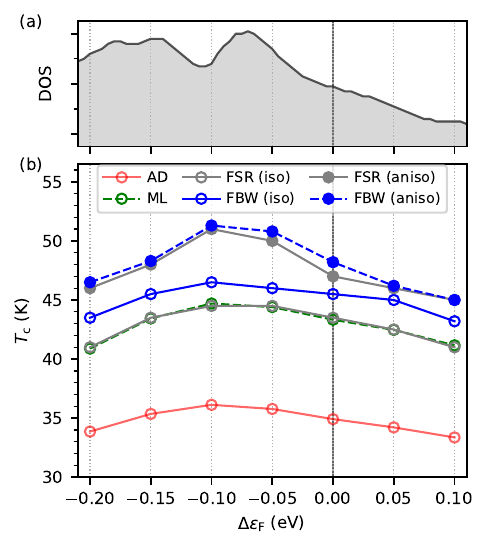}
    \caption{(a) Total DOS in the energy range of $-$0.2 to 0.1~eV around the Fermi level, and (b) superconducting critical temperature $T_{\rm c}$ as a function of Fermi level shift $\Delta \varepsilon_{\rm F}$ obtained with the AD formula, the ML model, and the isotropic and anisotropic ME equations in the FSR and FBW approaches with a $\mu^*=0.1$ for oS20-NaC$_4$ at 10~GPa.} 
    \label{fig:Tc-ef}
\end{figure}

Finally, we explored the effect of electron and hole doping on the $T_{\rm c}$ in the light of the vHS peak present in the DOS of oS20-NaC$_4$ near the Fermi level (see Fig.~\ref{fig:Tc-ef}(a)). Previous theoretical studies on H$_3$S have shown an enhancement in $T_{\rm c}$ when the Fermi level approaches the maximum of the vHS peak ~\cite{Sano2016,Quan2016,Lucrezi2024}. To investigate the dependence of $T_{\rm c}$ with doping, we shifted the Fermi level by $\Delta \varepsilon_{\rm F}$ in the $-$0.2~eV to 0.1~eV range in increments of 0.05~eV, corresponding to changes in the electron number of $\sim$0.076. Figure~\ref{fig:Tc-ef}(b) shows the $T_{\rm c}$ values obtained by solving the isotropic and anisotropic ME equations within the FSR and FBW approaches, and those obtained with the semi-empirical AD and ML formulas. Overall, all treatments produce a similar trend for the variation of $T_{\rm c}$ with doping. In particular, the $T_{\rm c}$ versus $\Delta \varepsilon_{\rm F}$ curve obtained with the anisotropic approach resembles the shape dependence of the DOS, the $T_{\rm c}$ reaching a maximum value at $\Delta \varepsilon_{\rm F} = -0.1$~eV when the Fermi level lies in very close proximity to the vHS peak. For comparison, the effect of doping on $T_{\rm c}$ is less pronounced when employing the isotropic ME treatment or the semi-empirical formulas, with the AD formula consistently underestimating the $T_{\rm c}$ values by about 7~K. 
%\vspace{5mm}

\section{\label{sec:summary}Summary}

We have explored the stability and superconductivity of Na-intercalated graphite compounds under moderate pressures using \textit{ab initio} methods. Through systematic screening, we identified new stable stoichiometries, Na$_3$C$_{10}$, NaC$_8$, NaC$_{10}$, and NaC$_{12}$, which redefine the previously established convex hulls up to 10~GPa. While some compounds may lack stability at zero temperature, they could be high-temperature ground states or form through cold compression of graphite. Our analysis of the nearly free electron states important for superconductivity in intercalated compounds narrows down the promising candidates to two. One of them, NaC$_4$ proposed previously by Hao {\it et al.}~\cite{Hao2023}, has been re-analyzed within the anisotropic Migdal-Eliashberg theory and predicted to be a two-gap superconductor with a $T_{\rm c}$ of 48 K, 24\% above the prior isotropic Eliashberg estimate. The other compound, Na$_3$C$_{10}$ introduced in the present study, appears to be prone to interlayer shifts and may develop domain walls to become dynamically stable. According to our estimates, it should have $T_{\rm c}$ at least comparable to that in NaC$_4$ at 10 GPa. The {\it ab initio} findings reveal the Na-C system may host viable compounds with unexpectedly high $T_{\rm c}$ for electron-doped materials.

\begin{acknowledgments}
S.B.M. and E.R.M. acknowledge support from the National Science Foundation (NSF) Award No. DMR-2035518. E.T.M. and A.N.K. acknowledge support from the NSF Award No. DMR-2320073. This study used the Frontera supercomputer at the Texas Advanced Computing Center through the Leadership Resource Allocation (LRAC) award DMR22004. Frontera is made possible by NSF award OAC-1818253~\cite{Frontera}. This study also used the Expanse system at the San Diego Supercomputer Center through allocation TG-DMR180071 from the Advanced Cyberinfrastructure Coordination Ecosystem: Services \& Support (ACCESS) program~\cite{ACCESS}, which is supported by NSF grants \#2138259, \#2138286, \#2138307, \#2137603, and \#2138296.
\end{acknowledgments}


%%%%%%%%%% Merge with supplemental materials %%%%%%%%%%

\widetext
\clearpage
\begin{center}
\textbf{\large Supplemental Material: \\ Stability-superconductivity map for compressed Na-intercalated graphite}
\end{center}

%%%%%%%%%% Merge with supplemental materials %%%%%%%%%%
%%%%%%%%%% Prefix a "S" to all equations, figures, tables and reset the counter %%%%%%%%%%

\makeatletter
\renewcommand{\theequation}{S\arabic{equation}}
\renewcommand{\thefigure}{S\arabic{figure}}
% \renewcommand{\bibnumfmt}[1]{[S#1]}
% \renewcommand{\citenumfont}[1]{S#1}
%%%%%%%%%% Prefix a "S" to all equations, figures, tables and reset the counter 
\setcounter{equation}{0}
\setcounter{figure}{0}
\setcounter{table}{0}
\setcounter{page}{1}
\renewcommand{\figurename}{FIG.}
\renewcommand{\thefigure}{S\arabic{figure}}
\renewcommand{\thetable}{S\arabic{table}}
\renewcommand{\thepage}{S\arabic{page}}

\begin{figure}[h]
    \centering
    \includegraphics[width=17cm]{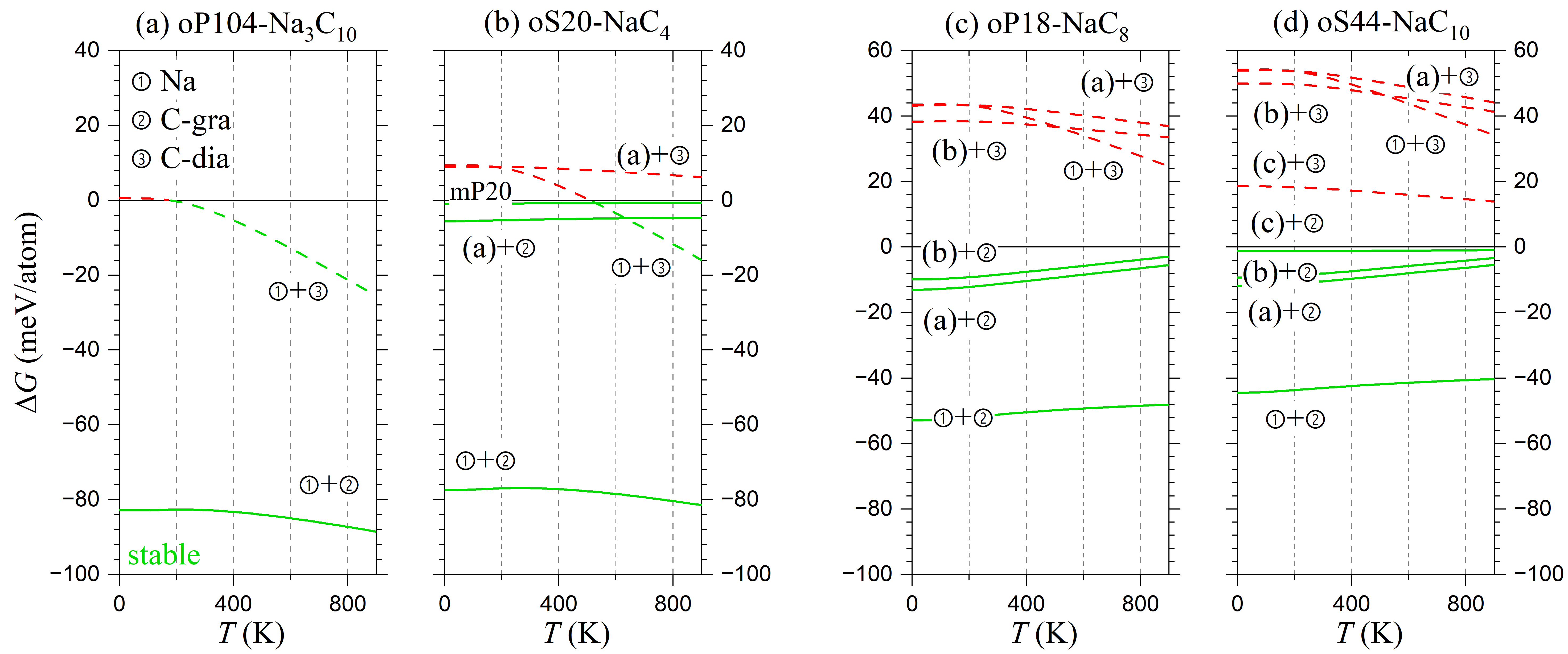}
    \caption{Relative Gibbs free energy at 10 GPa as a function of temperature for select Na-C phases calculated with the optB86b-vdW functional. Each line shows relative stability with respect to a combination of known (circled numbers) or proposed (bracketed letters) phases. The dashed lines correspond to phase combinations containing the diamond ground state at this pressure, which we argue to be not relevant for Na intercalation of graphite via cold compression. Since all solid lines measuring the relative stability with respect to graphite are below zero (shown in green), all four proposed phases are expected to be synthesizable in the considered temperature range. In addition, oP104-Na$_3$C$_{10}$ is globally stable above about 200 K.}
    \label{S1}
\end{figure}

\begin{figure}[h]
    \centering
    \includegraphics[width=17cm]{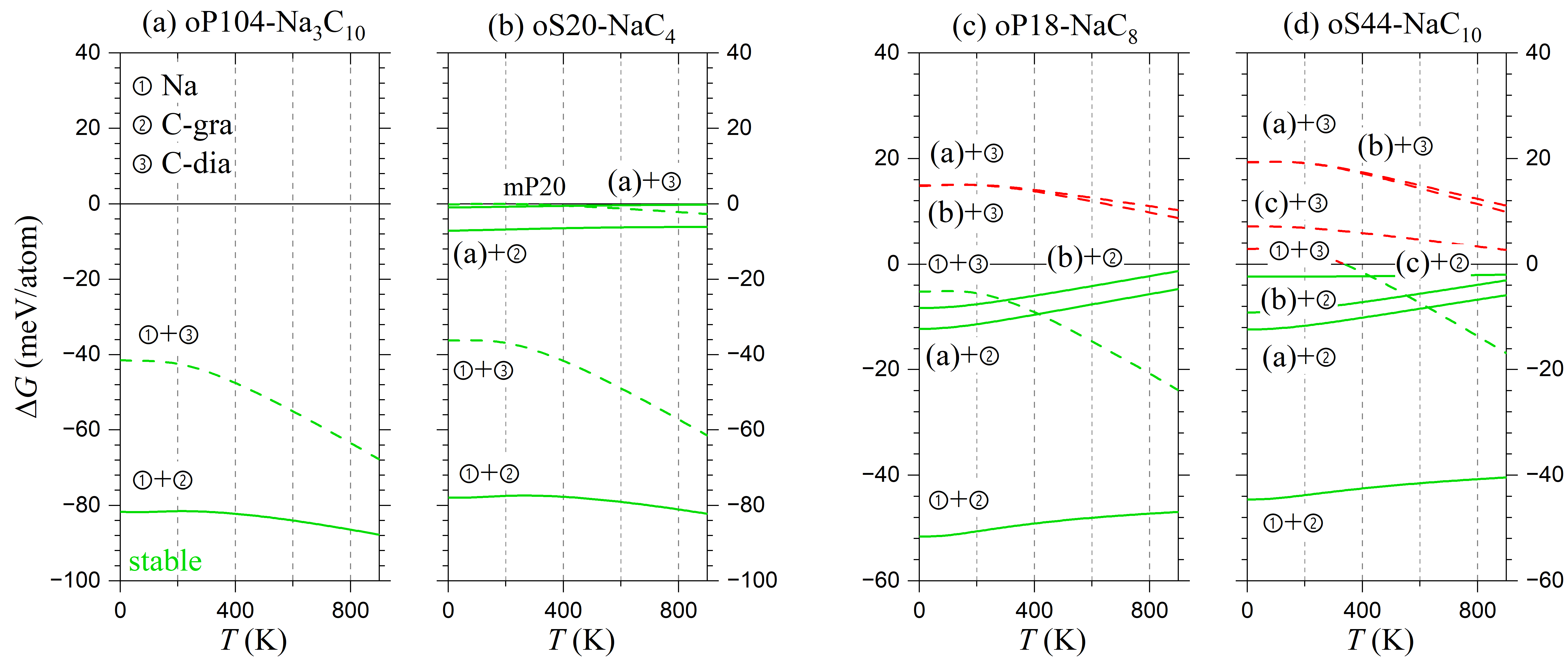}
    \caption{Relative Gibbs free energy at 10 GPa as a function of temperature for select Na-C phases calculated with the optB88-vdW functional (see Fig. S1 caption for further details). All four phases appear synthesizable in the considered temperature range if graphite is used as a starting material, with two of them, oP104-Na$_3$C$_{10}$ and oS20-NaC$_4$, being globally stable with respect to diamond.}
    \label{S2}
\end{figure}

\begin{figure}[h]
    \centering
    \includegraphics[width=12cm]{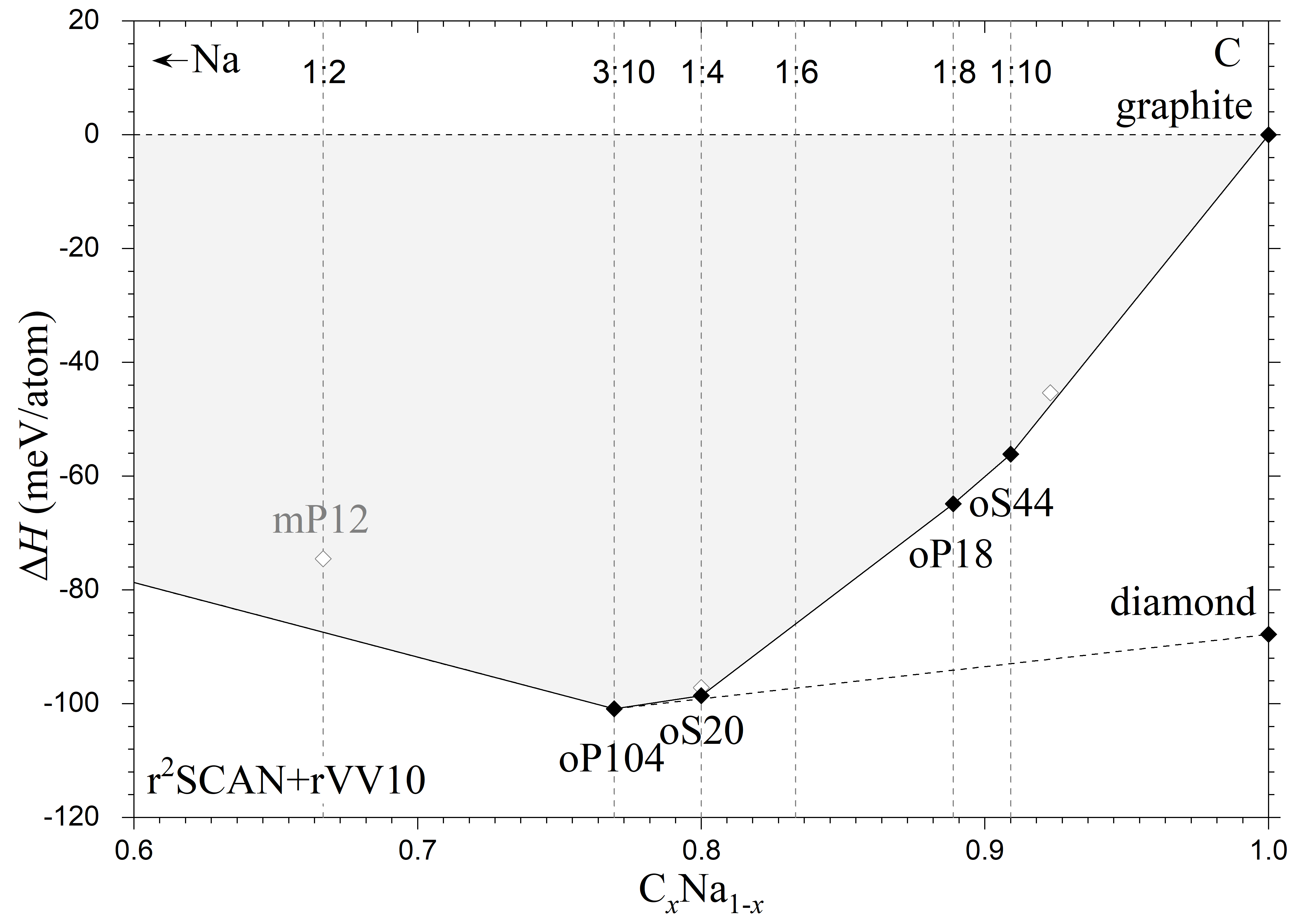}
    \caption{Stability of C-Na phases at 10 GPa and 0 K calculated with the r$^2$SCAN+rVV10 functional. The global (local) convex hulls are denoted with solid (dashed) lines. The set of four locally stable phases is consistent with those obtained with the optB88-vdW and optB86b-vdW functionals (see Fig. 2 in the manuscript).}
    \label{S3}
\end{figure}

\newpage

\begin{figure}[h]
    \centering
    \includegraphics[width=14cm]{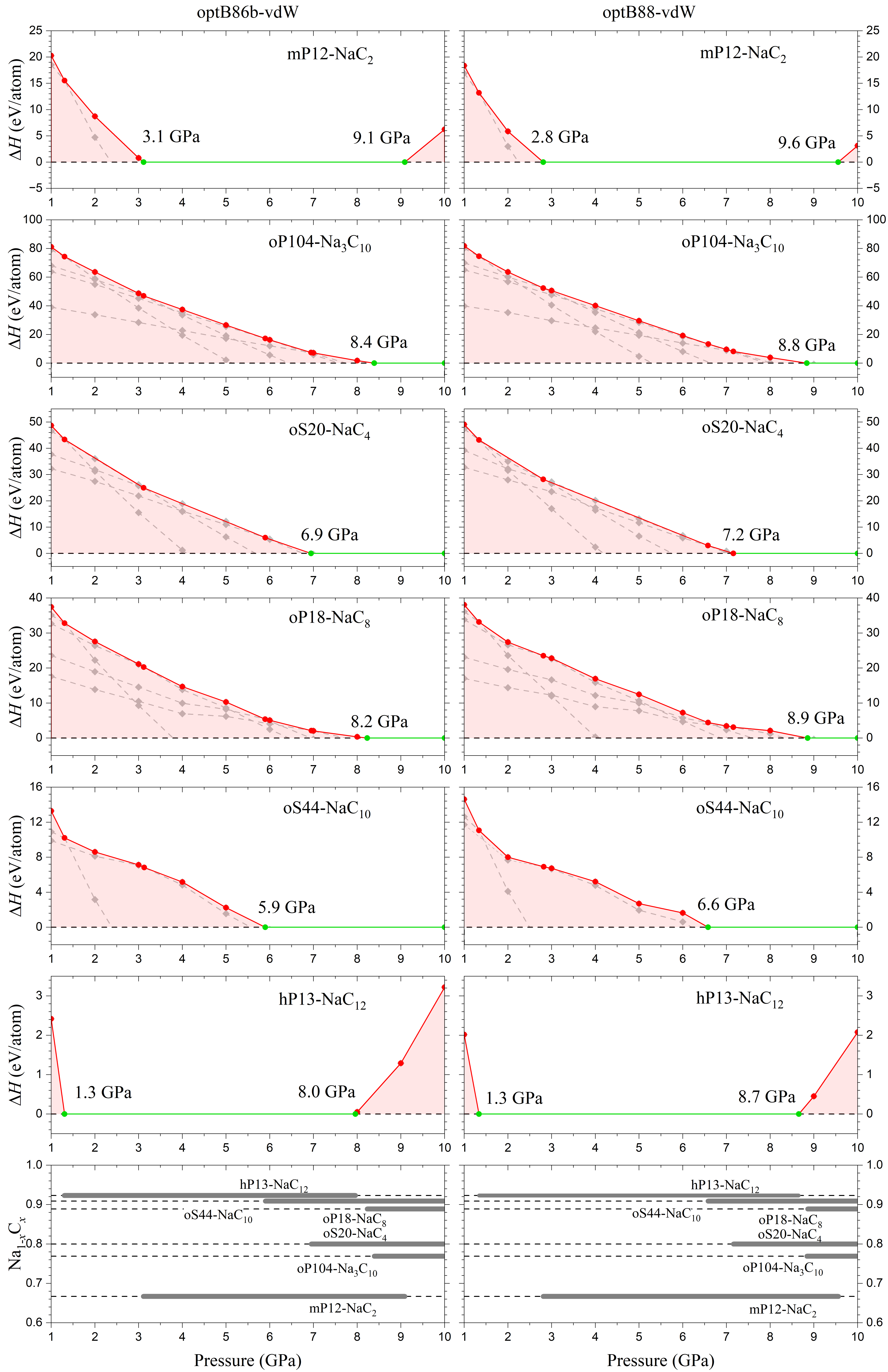}
    \caption{Stability of possible layered Na-C phases calculated at 0 K with the optB86b-vdW functional (left panels) and the optB88-vdW functional (right panels). The top six sets of panels illustrate the distance to the convex hull for each viable Na-C phase in the 1$-$10 GPa range. The red points with positive relative enthalpy values indicate phase metastability, while green points mark critical pressures at which the phase becomes or stops being a part of the convex hull. The two bottom panels summarize the results for stability ranges in the two approximations.}
    \label{S4}
\end{figure}

\begin{figure}[h]
    \centering
    \includegraphics[scale=0.60]{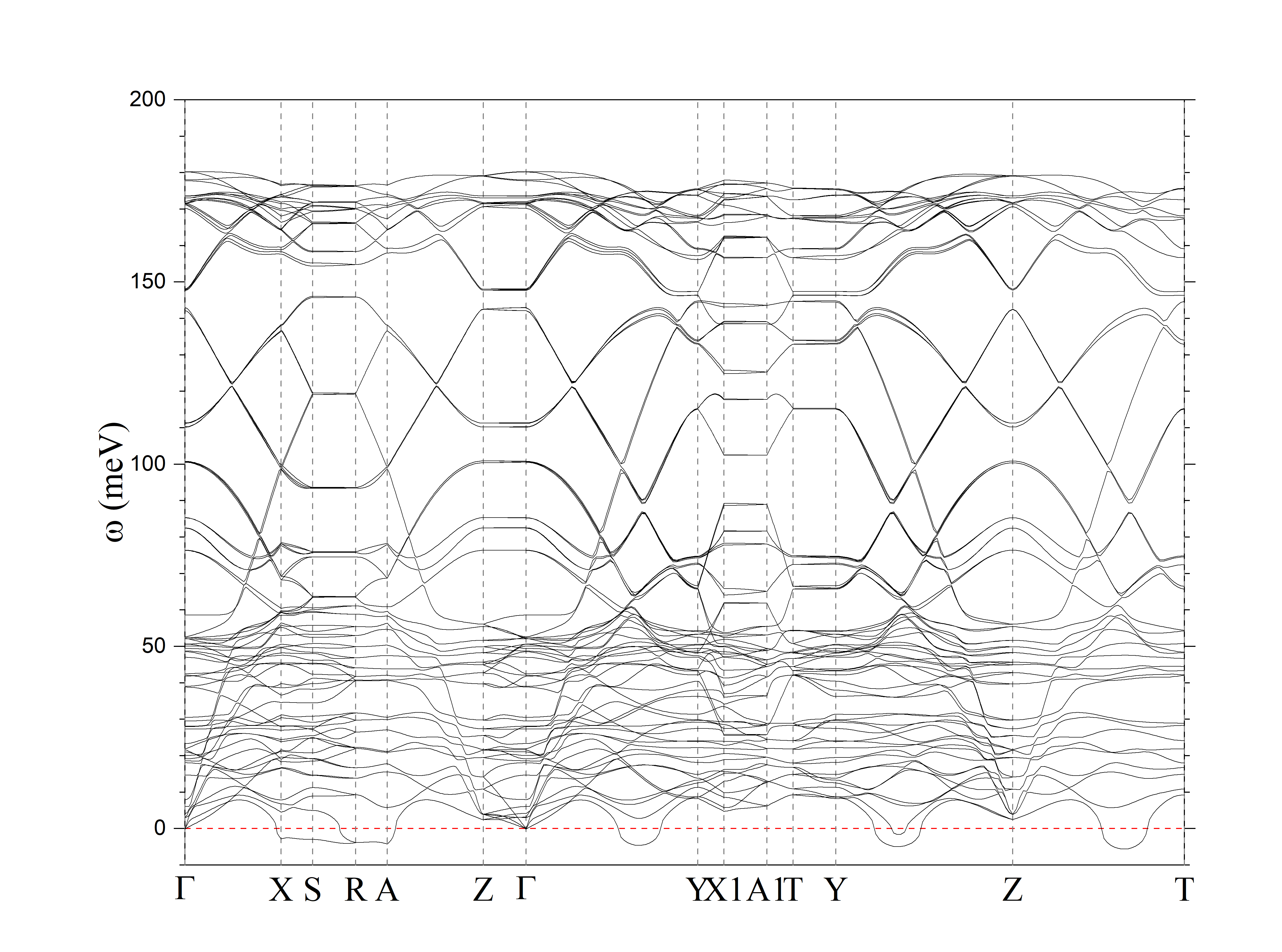}
    \caption{Phonon dispersion for oS52-Na$_3$C$_{10}$ ($Cmcm$) at 10~GPa calculated with the frozen phonon method using {\small PHONOPY} and {\small VASP}. Atoms in the $2\times 2\times 2$ expansion of the primitive unit cell were displaced by 0.04 \AA. Static calculations were performed with the optB86b-vdW functional and a $6\times 6\times 6$ $k$-point mesh.}
    \label{fig:076-phband}
\end{figure}

\begin{figure}[h]
    \centering
    \includegraphics[scale=0.60]{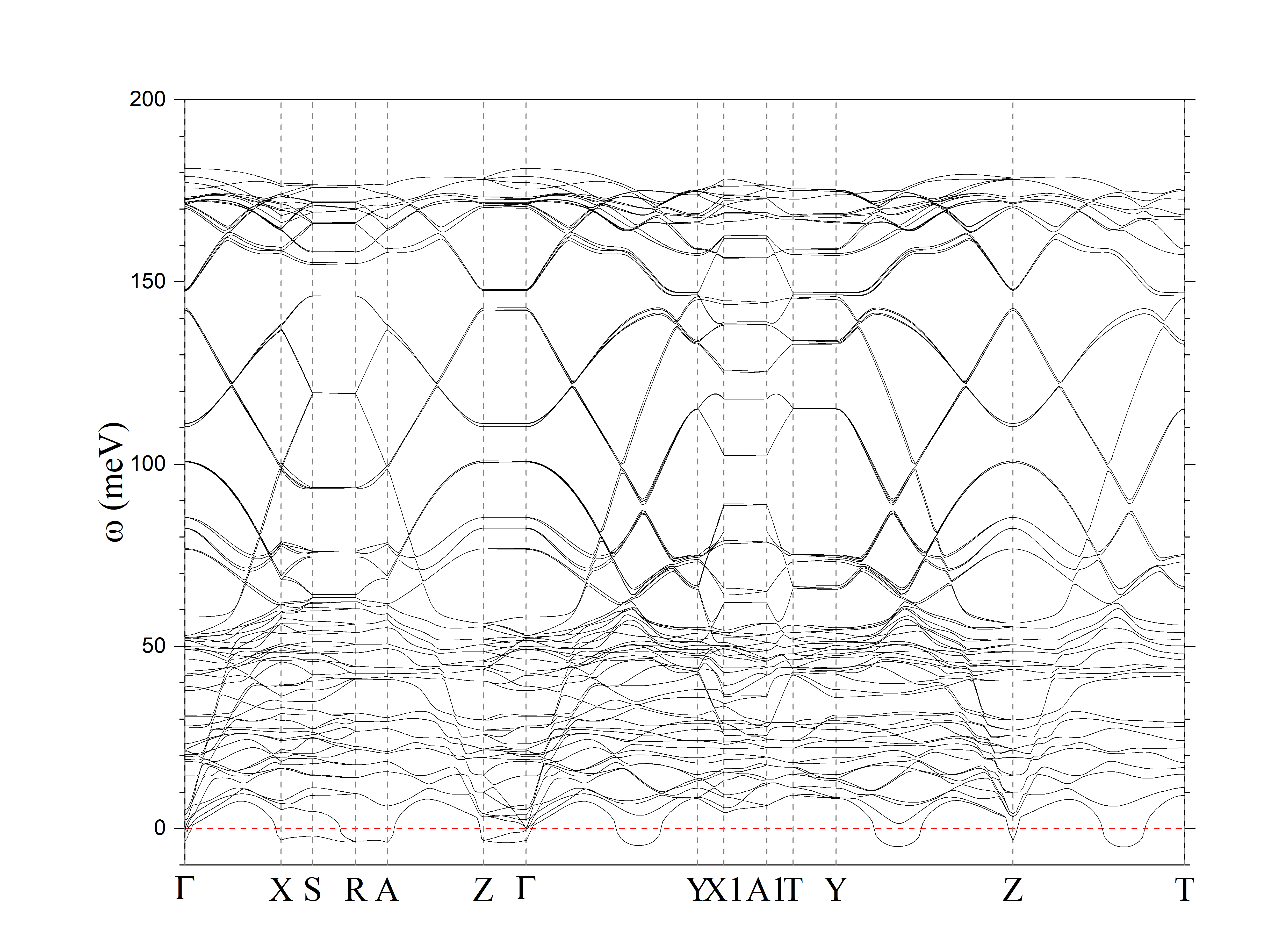}
    \caption{Phonon dispersion for oS52-Na$_3$C$_{10}$ ($C222_1$) at 10~GPa calculated with the frozen phonon method using {\small PHONOPY} and {\small VASP}. Atoms in the $2\times 2\times 2$ expansion of the primitive unit cell were displaced by 0.04 \AA. Static calculations were performed with the optB86b-vdW functional and a $6\times 6\times 6$ $k$-point mesh.}
    \label{fig:415-phband}
\end{figure}

\begin{figure}[h]
    \centering
    \includegraphics[scale=0.60]{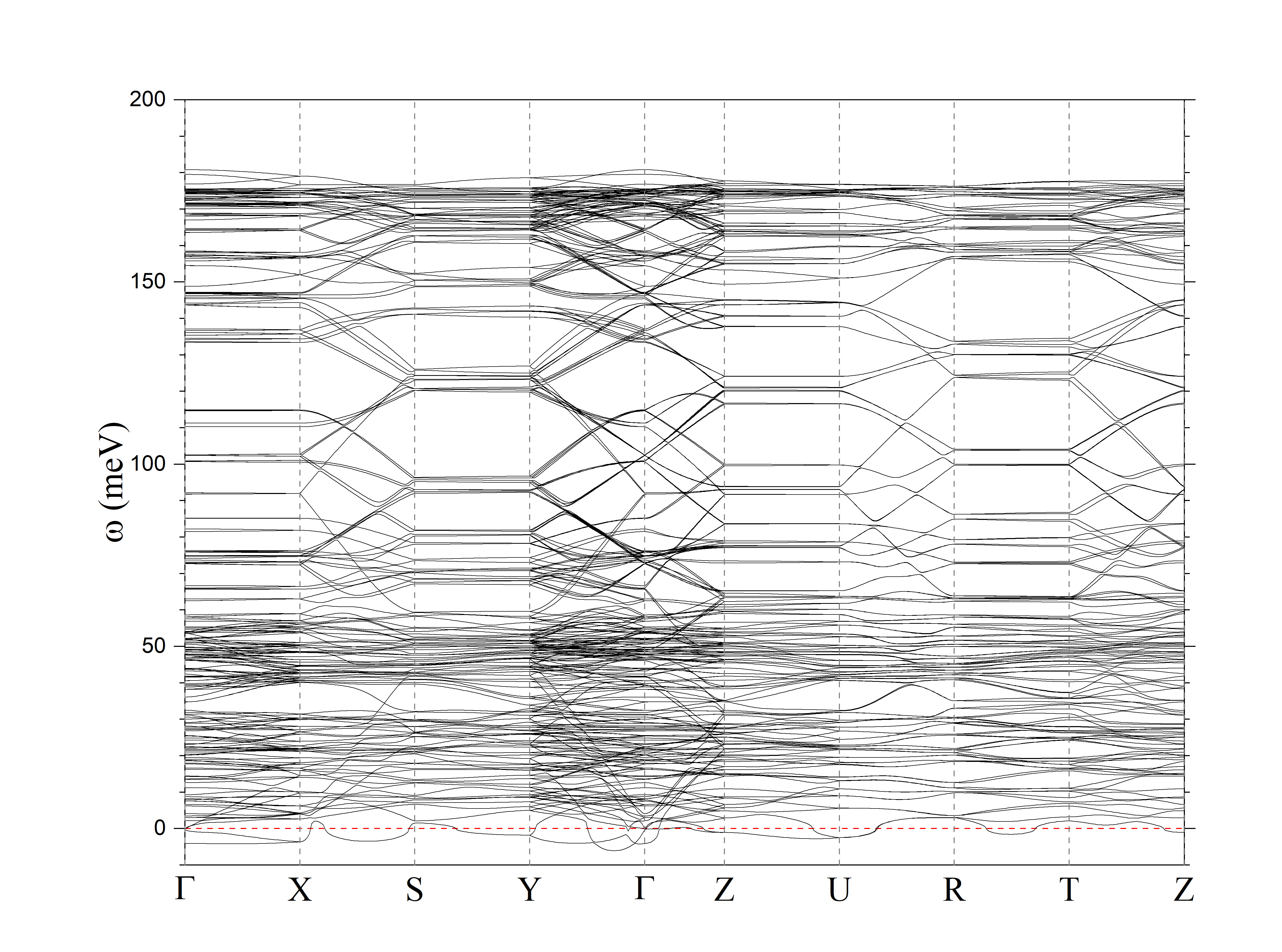}
    \caption{Phonon dispersion for oP104-Na$_3$C$_{10}$ at 10~GPa calculated with the frozen phonon method using {\small PHONOPY} and {\small VASP}. Atoms in the conventional unit cell were displaced by 0.04 \AA. Static calculations were performed with the optB86b-vdW functional and a $6\times 6\times 4$ $k$-point mesh. Note that lattice vectors in this orthorhombic unit cell are ordered by length and the similarity of the out-of-plane $a=8.6627$~\AA\ and in-plane $b=8.6645$~\AA\ is accidental. The reverse order of $a$ and $b$ in the unit cell optimized with Quantum ESPRESSO in Fig. \textcolor{blue}{S8} explains the swap of the corresponding segments in the $q$-point path in the Brillouin zone.}
    \label{fig:oP104-band}
\end{figure}

\begin{figure}[h]
    \centering
    \includegraphics[scale=0.99]{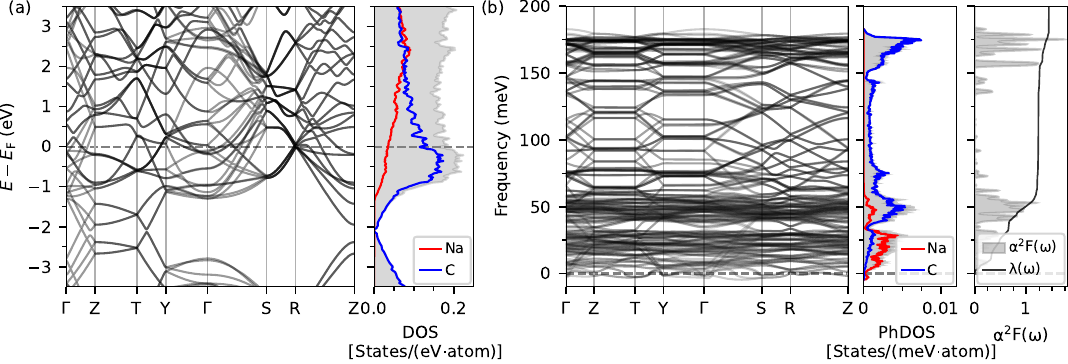}
    \caption{(a) Electronic band structure and density of states (DOS) and (b) phonon dispersion, phonon DOS (PhDOS), Eliashberg spectral function $\alpha^2 F(\omega)$, and integrated electron-phonon coupling strength $\lambda^{\rm QE}$ for oP104-Na$_3$C$_{10}$ at 10~GPa calculated with Quantum {\small ESPRESSO}. The linear response phonon calculations were only performed at the $\Gamma$-point, and the $\alpha^2F(\omega)$ and $\lambda^{\rm QE}$ were estimated by considering only the real phonon frequencies.} 
    \label{fig:structs-ph-a2f}
\end{figure}

\begin{figure*}[!hbt]
    \centering
    \includegraphics[scale=0.89]{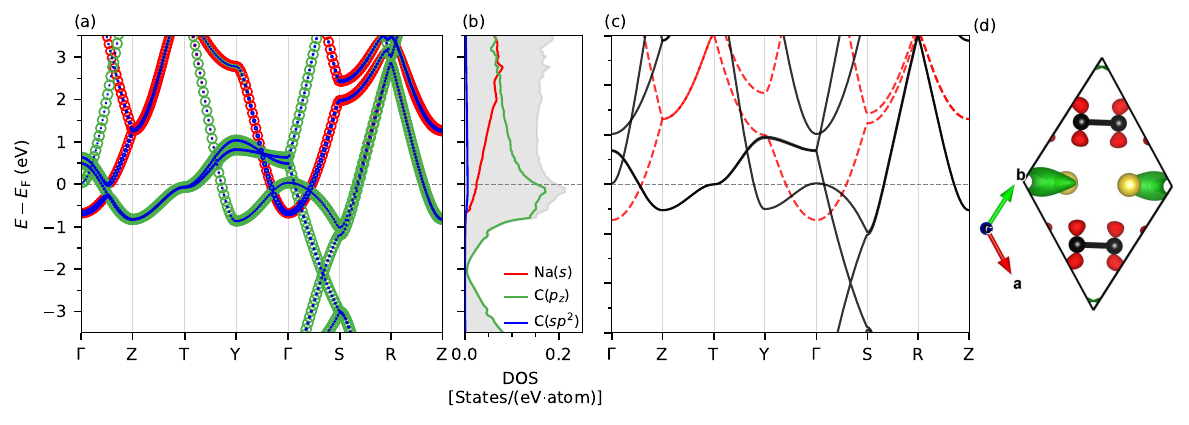}
    \caption{(a) Electronic band structure and (b) density of states (DOS) with orbital decomposition for oS20-NaC$_4$ at 10~GPa. The size of the symbols in (a) is proportional to the contribution of each orbital character. In (b), the total DOS is represented by the gray shaded region, and the projected DOS corresponding to Na-$s$, C-$p_z$, and C-$sp^2$ orbitals is shown with red, green, and blue lines, respectively. (c) Electronic band structure of C$_4$ structure (solid black) and Na structure (dashed red) with the same lattice parameters as for oS20-NaC$_4$. In (c), C$_4$ and Na bands are shifted downward and upward by $\sim$ 2~eV to match the Fermi level of NaC$_4$. (d) Charge density difference calculated as $\Delta \rho(r) = \rho_{\rm NaC_4}(r) - \rho_{\rm C_4}(r)- \rho_{\rm Na}(r)$ and plotted with VESTA. The red and green lobes represent the charge accumulation and depletion regions, respectively, with an isosurface value set to $7~\times 10^{-2}\, e/$\AA$^3$. }
    \label{fig:nac4-ele-band}
\end{figure*}

\begin{figure}
    \centering
    \includegraphics[scale=0.1]{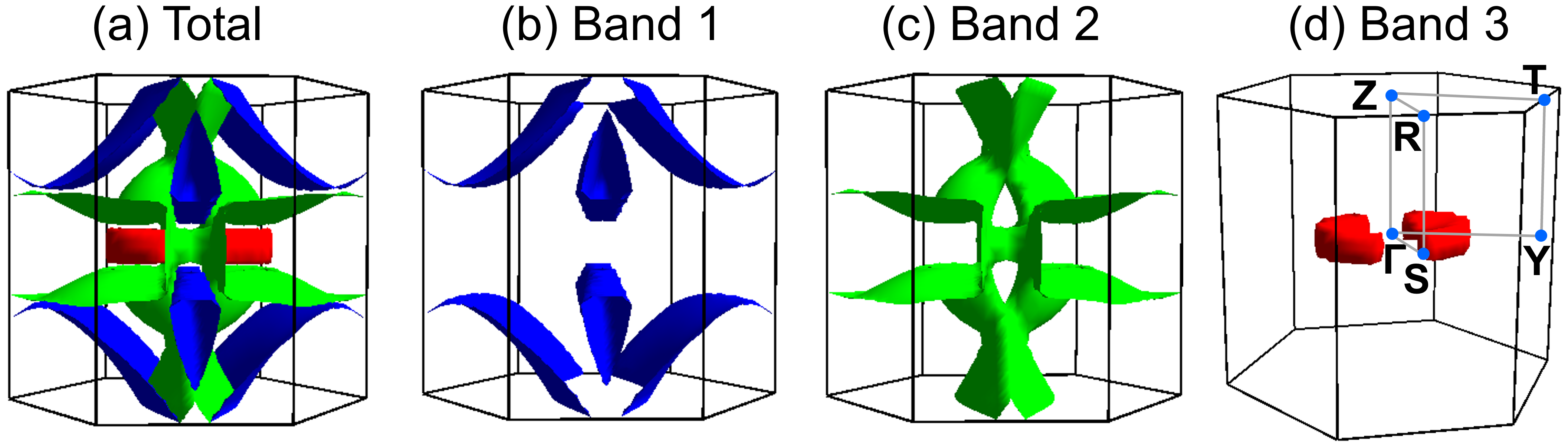}
    \caption{(a) Fermi surface (FS) in the first Brillouin zone, and (b$-$d) band decomposed FS corresponding to the three bands crossing the Fermi level for oS20-NaC$_4$ at 10~GPa. The plots were generated with FermiSurfer.}
    \label{fig:fs}
\end{figure}

\begin{figure*}[!hbt]
    \centering
    \includegraphics[scale=0.9]{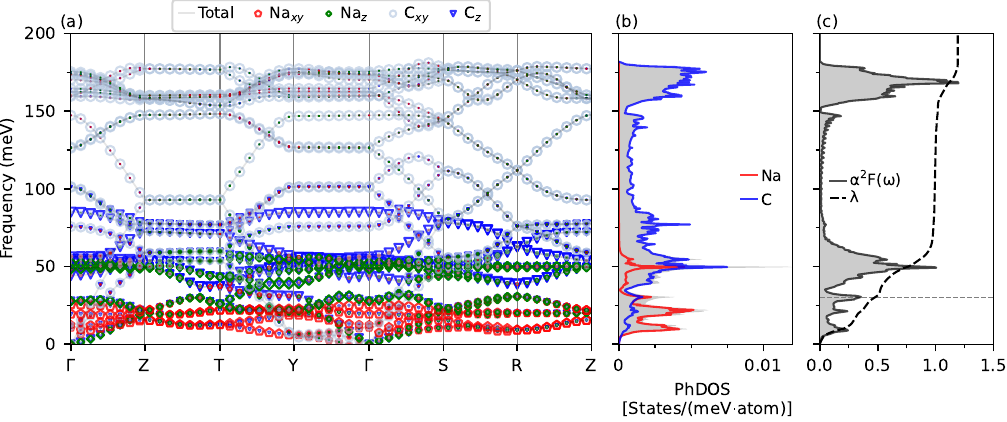}
    \caption{(a) Phonon band structure with decomposition into in-plane and out-of-plane Na and C vibrations, (b) total phonon density of states (PhDOS) shown as the gray-shaded region along with atomic-projected PhDOS, and (c) isotropic Eliashberg spectral function $\alpha^2F(\omega)$ and cumulative electron-phonon coupling strength $\lambda(\omega)$ for oS20-NaC$_4$ at 10~GPa. In (a), the size of the symbols is proportional to the amount of vibration. The results plotted in (a) and (b) were obtained with Quantum {\small ESPRESSO}, and those in (c) with EPW.}
    \label{fig:ph-band}
\end{figure*}

\begin{figure}[h]
    \centering
    \includegraphics[scale=0.22]{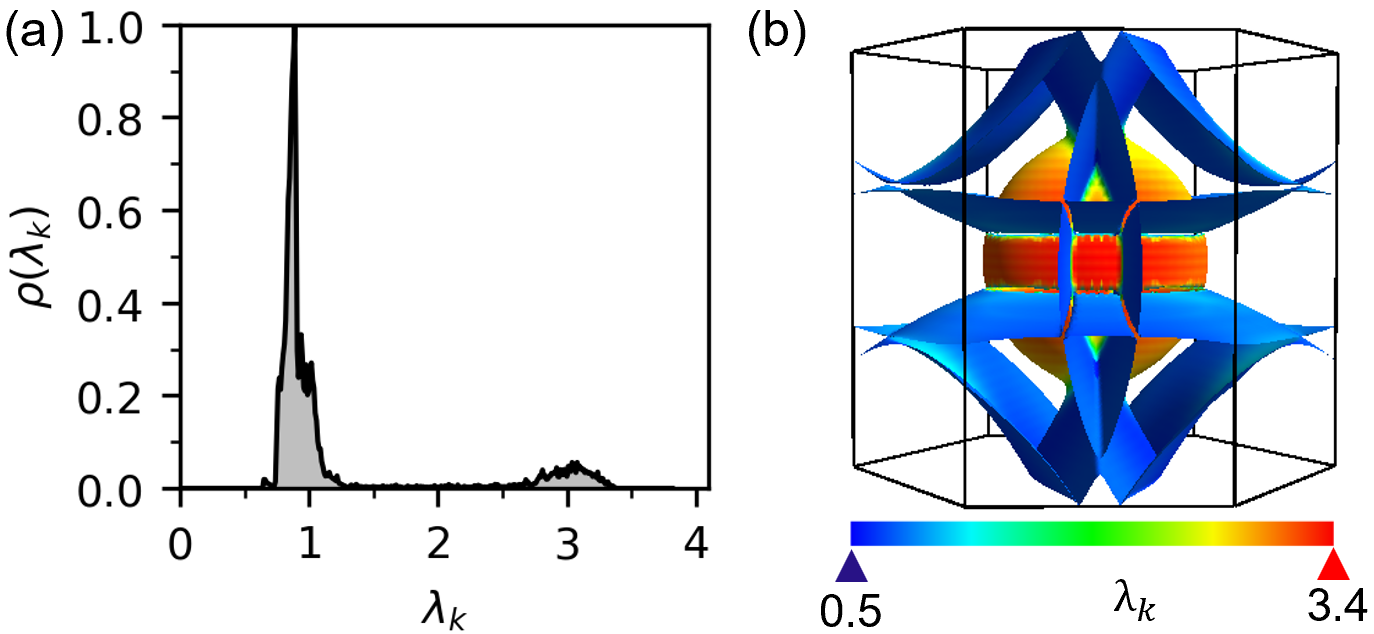}
    \caption{(a) Distribution of the electron-phonon coupling strength $\lambda_{\bm{k}}$, and (b) momentum-resolved electron-phonon coupling $\lambda_{\bm{k}}$ on the Fermi surface for oS20-NaC$_4$ at 10~GPa.} 
    \label{fig:lamda-aniso}
\end{figure}

\clearpage

%123
\begin{lstlisting}[caption={CIF file for mP12-NaC$_2$ containing the VASP optimized lattice constants and atomic positions at 5 GPa.},captionpos=t,frame=single]
_symmetry_Int_Tables_number 11
_cell_length_a    4.9385647015969827
_cell_length_b    4.2789321171193713
_cell_length_c    6.7852162684099380
_cell_angle_alpha 90.0000000000000000
_cell_angle_beta  96.7534216735655974
_cell_angle_gamma 90.0000000000000000
_symmetry_Int_Tables_number 11
_chemical_formula_sum
'C Na '
loop_
_atom_site_label
_atom_site_type_symbol
_atom_site_site_symmetry_multiplicity
_atom_site_occupancy wyckoff
_atom_site_fract_x
_atom_site_fract_y
_atom_site_fract_z
C1 C    4 f  0.3738522465525522  0.9170128071125633  0.9915317221880430
C2 C    4 f  0.8755110898962359  0.9166666666666666  0.0032723074604634
Na1 Na    2 e  0.5209942629097670  0.2500000000000000  0.6611028305808305
Na2 Na    2 e  0.0333352330481349  0.7500000000000000  0.6676964553998347
#End
\end{lstlisting}

\clearpage

%076
\begin{lstlisting}[caption={CIF file for oS52-Na$_3$C$_{10}$ ($Cmcm$) containing the VASP optimized lattice constants and atomic positions at 10 GPa.},captionpos=t,frame=single]
_symmetry_Int_Tables_number 63
_cell_length_a    4.3308489584442942
_cell_length_b    12.4624930111442254
_cell_length_c    8.6752910540271539
_cell_angle_alpha 90.0000000000000000
_cell_angle_beta  90.0000000000000000
_cell_angle_gamma 90.0000000000000000
_symmetry_Int_Tables_number 63
_chemical_formula_sum
'C Na '
loop_
_atom_site_label
_atom_site_type_symbol
_atom_site_site_symmetry_multiplicity
_atom_site_occupancy wyckoff
_atom_site_fract_x
_atom_site_fract_y
_atom_site_fract_z
C1 C   16 h  0.8329904911974388  0.2001191736615588  0.4924281340166700
C2 C   16 h  0.3333333333333333  0.1000192751768089  0.0131698907471501
C3 C    8 e  0.1671755795227128  0.0000000000000000  0.0000000000000000
Na1 Na    4 c  0.0000000000000000  0.4342389766509883  0.2500000000000000
Na2 Na    4 c  0.0000000000000000  0.2564419263529202  0.7500000000000000
Na3 Na    4 c  0.0000000000000000  0.0919159442533302  0.2500000000000000
#End
\end{lstlisting}

%415
\begin{lstlisting}[caption={CIF file for oS52-Na$_3$C$_{10}$ ($C222_1$) containing the VASP optimized lattice constants and atomic positions at 10 GPa.},captionpos=t,frame=single]
_symmetry_Int_Tables_number 20
_cell_length_a    4.3302272757988716
_cell_length_b    12.4591708232311316
_cell_length_c    8.6730514489049142
_cell_angle_alpha 90.0000000000000000
_cell_angle_beta  90.0000000000000000
_cell_angle_gamma 90.0000000000000000
_symmetry_Int_Tables_number 20
_chemical_formula_sum
'C Na '
loop_
_atom_site_label
_atom_site_type_symbol
_atom_site_site_symmetry_multiplicity
_atom_site_occupancy wyckoff
_atom_site_fract_x
_atom_site_fract_y
_atom_site_fract_z
C1 C    8 c  0.8568009110072801  0.2003733256250904  0.4905135003996576
C2 C    8 c  0.3094567742800791  0.1001450689296917  0.0095053905966793
C3 C    4 a  0.1433842728892287  0.0000000000000000  0.0000000000000000
C4 C    8 c  0.6906653888959298  0.3001513881393549  0.5045861333095161
C5 C    8 c  0.1430804899867759  0.4001926934717837  0.9829425619190530
C6 C    4 a  0.8092422335363665  0.0000000000000000  0.0000000000000000
Na1 Na    4 b  0.0000000000000000  0.2587406547859601  0.7500000000000000
Na2 Na    4 b  0.0000000000000000  0.0902370332642235  0.2500000000000000
Na3 Na    4 b  0.5000000000000000  0.0695580858043456  0.7500000000000000
#End
\end{lstlisting}

\clearpage

%417
\begin{lstlisting}[caption={CIF file for oP104-Na$_3$C$_{10}$ containing the VASP optimized lattice constants and atomic positions at 10 GPa.},captionpos=t,frame=single]
_symmetry_Int_Tables_number 19
_cell_length_a    8.6627096907197174
_cell_length_b    8.6644816709095789
_cell_length_c    12.4570727597858930
_cell_angle_alpha 90.0000000000000000
_cell_angle_beta  90.0000000000000000
_cell_angle_gamma 90.0000000000000000
_symmetry_Int_Tables_number 19
_chemical_formula_sum
'C Na '
loop_
_atom_site_label
_atom_site_type_symbol
_atom_site_site_symmetry_multiplicity
_atom_site_occupancy wyckoff
_atom_site_fract_x
_atom_site_fract_y
_atom_site_fract_z
C1 C    4 a  0.0089769440136607  0.2005037301172762  0.9502844373296782
C2 C    4 a  0.4889723001143621  0.4661238549051346  0.8501092096543489
C3 C    4 a  0.5000000000000000  0.5493895059355721  0.7500000000000000
C4 C    4 a  0.9943327955796731  0.2838085626627843  0.0500860064139844
C5 C    4 a  0.5164850584687599  0.5495756853819741  0.1501607188076123
C6 C    4 a  0.5000000000000000  0.2162895280911100  0.7500000000000000
C7 C    4 a  0.5089551700446648  0.5495042345978048  0.5497369227922206
C8 C    4 a  0.9884799885466791  0.2838708977002468  0.6499225249215074
C9 C    4 a  0.9997685780013086  0.2006094410967164  0.7500000000000000
C10 C    4 a  0.4942931621564098  0.4661940575904352  0.4499390086830601
C11 C    4 a  0.0163740215540429  0.2004318618675639  0.3498563674054722
C12 C    4 a  0.0000000000000000  0.5337101835865851  0.7500000000000000
C13 C    4 a  0.9910799716819199  0.2005054322901428  0.5497285572808863
C14 C    4 a  0.5115598530096078  0.4661257309068317  0.6499318914036536
C15 C    4 a  0.0055537160765198  0.2838038672070055  0.4499259547892009
C16 C    4 a  0.4839766594770580  0.5495866317266332  0.3498744603162365
C17 C    4 a  0.4910906792366276  0.5494910142600630  0.9502973551977975
C18 C    4 a  0.0110767079808767  0.2838780565145170  0.8500996376184632
C19 C    4 a  0.5055468857352067  0.4661886503444349  0.0500947541777634
C20 C    4 a  0.9838913581277424  0.2004067856847412  0.1501408462742547
Na1 Na    4 a  0.2500000000000000  0.3751656366023609  0.6787822131660448
Na2 Na    4 a  0.7500000000000000  0.6244618188996894  0.0098059806446430
Na3 Na    4 a  0.2500000000000000  0.6255374714362886  0.4904625713022241
Na4 Na    4 a  0.7500000000000000  0.3744437684065572  0.1603035422400512
Na5 Na    4 a  0.2500000000000000  0.3755451728489588  0.3401661341425566
Na6 Na    4 a  0.7500000000000000  0.6251419523376142  0.3192047953244241
#End
\end{lstlisting}

\clearpage

%417-QE optimized structure
\begin{lstlisting}[caption={CIF file for oP104-Na$_3$C$_{10}$ containing the Quantum {\small ESPRESSO} optimized lattice constants and atomic positions at 10 GPa.},captionpos=t,frame=single]
_symmetry_Int_Tables_number 19
_cell_length_a    8.6669530868999995
_cell_length_b    8.6752223969000006
_cell_length_c    12.4665222167999996
_cell_angle_alpha 90.0000000000000000
_cell_angle_beta  90.0000000000000000
_cell_angle_gamma 90.0000000000000000
_symmetry_Int_Tables_number 19
_chemical_formula_sum
'C Na '
loop_
_atom_site_label
_atom_site_type_symbol
_atom_site_site_symmetry_multiplicity
_atom_site_occupancy wyckoff
_atom_site_fract_x
_atom_site_fract_y
_atom_site_fract_z
C1 C    4 a  0.4507236480000000  0.2413245620000000  0.2002537025000000
C2 C    4 a  0.7161226870000000  0.7610236880000001  0.1000912565000000
C3 C    4 a  0.7991788980000000  0.7500000000000000  0.0000000000000000
C4 C    4 a  0.5338008400000001  0.2556376930000002  0.3000730205000001
C5 C    4 a  0.7993377450000000  0.7338476060000001  0.4001396815000001
C6 C    4 a  0.4662631750000000  0.7500000000000000  0.0000000000000000
C7 C    4 a  0.7992854120000000  0.7413208360000001  0.7997665335000000
C8 C    4 a  0.5338787440000000  0.2615109260000001  0.8999348805000000
C9 C    4 a  0.4508216080000000  0.2500000000000000  0.0000000000000000
C10 C    4 a  0.7162031530000000  0.7556124210000001  0.6999438925000000
C11 C    4 a  0.4506610040000000  0.2339449550000001  0.5998697805000001
C12 C    4 a  0.7837345600000001  0.2500000000000000  0.0000000000000000
C13 C    4 a  0.4507303540000001  0.2586128890000001  0.7997603935000001
C14 C    4 a  0.7161152360000002  0.7384610060000001  0.8999432845000001
C15 C    4 a  0.5337959530000000  0.2445494410000002  0.6999299455000001
C16 C    4 a  0.7993500830000000  0.7656854870000001  0.5998865295000001
C17 C    4 a  0.7992800470000000  0.7585774540000001  0.2002647895000000
C18 C    4 a  0.5338860750000000  0.2389359800000002  0.1000817115000000
C19 C    4 a  0.7161939740000000  0.7444899560000000  0.3000872955000001
C20 C    4 a  0.4506293540000000  0.2657527510000002  0.4001268375000001
Na1 Na    4 a  0.8744547370000000  0.9999537350000001  0.0901991205000000
Na2 Na    4 a  0.6249402170000000  0.5000000000000000  0.0693910415000000
Na3 Na    4 a  0.8755095600000000  0.9999712590000001  0.7408528845000001
Na4 Na    4 a  0.6244340540000000  0.5000000000000000  0.4101835415000000
Na5 Na    4 a  0.6250827310000000  0.0000000000000000  0.9286509085000001
Na6 Na    4 a  0.8745149970000001  0.5000000000000000  0.2596195865000001
#End
\end{lstlisting}

\clearpage

%132
\begin{lstlisting}[caption={CIF file for oS20-NaC$_{4}$ containing the VASP optimized lattice constants and atomic positions at 10 GPa.},captionpos=t,frame=single]
_symmetry_Int_Tables_number 63
_cell_length_a    8.4416196582911862
_cell_length_b    4.9766024157034465
_cell_length_c    4.3180055439772715
_cell_angle_alpha 90.0000000000000000
_cell_angle_beta  90.0000000000000000
_cell_angle_gamma 90.0000000000000000
_symmetry_Int_Tables_number 63
_chemical_formula_sum
'C Na '
loop_
_atom_site_label
_atom_site_type_symbol
_atom_site_site_symmetry_multiplicity
_atom_site_occupancy wyckoff
_atom_site_fract_x
_atom_site_fract_y
_atom_site_fract_z
C1 C   16 h  0.7430429715251350  0.1250633379757815  0.5833333333333334
Na1 Na    4 c  0.5000000000000000  0.7885882983452959  0.7500000000000000
#End
\end{lstlisting}

%259
\begin{lstlisting}[caption={CIF file for mP20-NaC$_{4}$ containing the VASP optimized lattice constants and atomic positions at 10 GPa.},captionpos=t,frame=single]
_symmetry_Int_Tables_number 11
_cell_length_a    4.9765419789501575
_cell_length_b    4.2990571945915708
_cell_length_c    8.6689322899801819
_cell_angle_alpha 90.0000000000000000
_cell_angle_beta  99.1740407688549794
_cell_angle_gamma 90.0000000000000000
_symmetry_Int_Tables_number 11
_chemical_formula_sum
'C Na '
loop_
_atom_site_label
_atom_site_type_symbol
_atom_site_site_symmetry_multiplicity
_atom_site_occupancy wyckoff
_atom_site_fract_x
_atom_site_fract_y
_atom_site_fract_z
C1 C    4 f  0.8749717507573713  0.9170693501666208  0.0013319747396636
C2 C    4 f  0.3724445126662310  0.9177921869421583  0.9883987677625825
C3 C    4 f  0.1276894349365955  0.4178826404499106  0.5118205685833459
C4 C    4 f  0.6249989366523643  0.4170243017585875  0.4988372940684156
Na1 Na    2 e  0.9834529715650895  0.2500000000000000  0.2493182820730058
Na2 Na    2 e  0.5180368802956501  0.7500000000000000  0.2503183999088704
#End
\end{lstlisting}

\clearpage

%011
\begin{lstlisting}[caption={CIF file for oP18-NaC$_{8}$ containing the VASP optimized lattice constants and atomic positions at 10 GPa.},captionpos=t,frame=single]
_symmetry_Int_Tables_number 51
_cell_length_a    4.2850729223012243
_cell_length_b    7.0397888933067252
_cell_length_c    4.9450845858175123
_cell_angle_alpha 90.0000000000000000
_cell_angle_beta  90.0000000000000000
_cell_angle_gamma 90.0000000000000000
_symmetry_Int_Tables_number 51
_chemical_formula_sum
'C Na '
loop_
_atom_site_label
_atom_site_type_symbol
_atom_site_site_symmetry_multiplicity
_atom_site_occupancy wyckoff
_atom_site_fract_x
_atom_site_fract_y
_atom_site_fract_z
C1 C    8 l  0.5833333333333334  0.8018946504893164  0.1243662737983722
C2 C    8 l  0.9166666666666666  0.7962948369854671  0.6253713887117981
Na1 Na    2 f  0.2500000000000000  0.5000000000000000  0.6987535822271693
#End
\end{lstlisting}

%006
\begin{lstlisting}[caption={CIF file for oS44-NaC$_{10}$ containing the VASP optimized lattice constants and atomic positions at 10 GPa.},captionpos=t,frame=single]
_symmetry_Int_Tables_number 65
_cell_length_a    4.2691888914443883
_cell_length_b    12.3407290006321055
_cell_length_c    7.0256501131042892
_cell_angle_alpha 90.0000000000000000
_cell_angle_beta  90.0000000000000000
_cell_angle_gamma 90.0000000000000000
_symmetry_Int_Tables_number 65
_chemical_formula_sum
'C Na '
loop_
_atom_site_label
_atom_site_type_symbol
_atom_site_site_symmetry_multiplicity
_atom_site_occupancy wyckoff
_atom_site_fract_x
_atom_site_fract_y
_atom_site_fract_z
C1 C   16 r  0.8330517804636273  0.2998826694963697  0.2986743462506494
C2 C   16 r  0.6661981370281538  0.3998344125430791  0.2940359521789800
C3 C    8 o  0.6673130245858871  0.0000000000000000  0.7079208385608736
Na1 Na    4 i  0.0000000000000000  0.3807638450300530  0.0000000000000000
#End
\end{lstlisting}

\clearpage

%004
\begin{lstlisting}[caption={CIF file for hP13-NaC$_{12}$ containing the VASP optimized lattice constants and atomic positions at 5 GPa.},captionpos=t,frame=single]
_symmetry_Int_Tables_number 191
_cell_length_a    4.2827286628374583
_cell_length_b    4.2827286628374583
_cell_length_c    7.2725399518830214
_cell_angle_alpha 90.0000000000000000
_cell_angle_beta  90.0000000000000000
_cell_angle_gamma 120.0000000000000142
_symmetry_Int_Tables_number 191
_chemical_formula_sum
'C Na '
loop_
_atom_site_label
_atom_site_type_symbol
_atom_site_site_symmetry_multiplicity
_atom_site_occupancy wyckoff
_atom_site_fract_x
_atom_site_fract_y
_atom_site_fract_z
C1 C   12 n  0.6666666666666666  0.0000000000000000  0.2063351542498993
Na1 Na    1 b  0.0000000000000000  0.0000000000000000  0.5000000000000000
#End
\end{lstlisting}

%\bibliography{pap}

\end{document}